\documentclass[journal,twoside]{IEEEtran}

\ifCLASSINFOpdf
\usepackage[pdftex]{graphicx}

\else

\fi

\usepackage[cmex10]{amsmath}
\usepackage{amssymb}   
\usepackage{amsxtra}
\usepackage{amscd}
\usepackage{amsthm}
\usepackage{textcomp}
\usepackage{graphicx}

\usepackage{balance}

\usepackage{array}  

\usepackage{multirow}  

\usepackage{amsmath}
\usepackage{cite}

\usepackage{url}

\usepackage{color,soul} 
\soulregister\cite7
\soulregister\ref7
\soulregister\pageref7

\begin{document}
	
	\title{{\huge Indoor and Outdoor Physical Channel Modeling and\\ Efficient Positioning for Reconfigurable Intelligent Surfaces in mmWave Bands}}

	
	%
	%
	%
	
	\author{Ertugrul~Basar,~\IEEEmembership{Senior Member,~IEEE}, Ibrahim~Yildirim,~\IEEEmembership{Graduate Student Member,~IEEE} and Fatih~Kilinc,~\IEEEmembership{Student Member,~IEEE} 

		\vspace*{-0.25cm}
		\thanks{E. Basar, I. Yildirim and F. Kilinc are with the Communications Research and Innovation Laboratory (CoreLab), Department of Electrical and Electronics Engineering, Ko\c{c} University, Sariyer 34450, Istanbul, Turkey. e-mail: ebasar@ku.edu.tr, fkilinc20@ku.edu.tr}
		\thanks{I. Yildirim is also with the Faculty of Electrical and Electronics Engineering, Istanbul Technical University, Istanbul 34469, Turkey. e-mail: yildirimib@itu.edu.tr}
		\thanks{This work was supported by the Scientific and Technological Research Council of Turkey (TUBITAK) under Grant 120E401.}
		\thanks{This paper was presented in part at 12th IEEE Latin-American Conference on Communications (IEEE LATINCOM 2020) and received the Best Paper Award \cite{SimRIS_latincom}.}
		\thanks{Codes available at https://corelab.ku.edu.tr/tools/SimRIS }
	}

	\maketitle

	\begin{abstract}
		Reconfigurable intelligent surface (RIS)-assisted communication appears as one of the potential enablers for sixth generation (6G) wireless networks by providing  a new degree of freedom in the system design to telecom operators. Particularly, RIS-empowered millimeter wave (mmWave) communication systems can be a remedy to provide broadband and ubiquitous connectivity. This paper aims to fill an important gap in the open literature by providing a physical, accurate, open-source, and widely applicable RIS channel model for mmWave frequencies. Our model is not only applicable in various indoor and outdoor environments but also includes the physical characteristics of wireless propagation in the presence of RISs by considering 5G radio channel conditions. Various deployment scenarios are presented for RISs and useful insights are provided for system designers from the perspective of potential RIS use-cases and their efficient positioning. The scenarios in which the use of an RIS makes a big difference or might not have a big impact on the communication system performance, are revealed. The open-source and comprehensive \textit{SimRIS Channel Simulator} is also introduced in this paper.
		
	\end{abstract}
	


	\IEEEpeerreviewmaketitle
	
	\section{Introduction}
	
	\IEEEPARstart{S}{ixth} generation (6G) wireless systems are expected to provide broadband connectivity by supporting new use-cases including extreme capacity and very high mobility, integrated with satellite networks and autonomous systems \cite{Rajatheva_6G}. However, these attractive features might only be possible with effective tools such as ultra massive multiple-input multiple-output (MIMO) systems, millimeter wave (mmWave) and TeraHertz (THz) communications, reconfigurable intelligent surfaces (RISs), cell-free networks, and integrated space and terrestrial networks. While 5G wireless networks are being introduced in various countries worldwide, wireless experts have already set their sights on 6G networking by starting active research on these interesting technologies \cite{Saad_2019}.
	
	RIS-empowered communication has received tremendous interest from the wireless research community due to its undeniable potential in extending the coverage, enhancing the link capacity, mitigating interference, deep fading, and Doppler effects, and increasing the physical layer (PHY) security \cite{Wu_2019,Basar_Access_2019,Basar_Doppler, yildirim2019propagation}. This can be accomplished by controlling the wireless propagation through unique electromagnetic (EM) functionalities provided by RISs. Numerous studies from the past two years have demonstrated that RISs, which are artificial, electronically controlled, 2D surfaces of EM material, can be effectively used to boost the performance of existing communication systems by exploiting the inherent randomness of wireless propagation \cite{Dardari, Renzo_JSAC}. Notable use-cases of RISs include energy efficient single/multi-user MIMO system designs with joint beamforming, PHY security systems, non-orthogonal multiple access schemes, index modulation systems, mmWave systems, vehicular/aerial networks, cognitive radio networks, wireless power transfer systems, posture recognition, and radio localization, and so on.   
	
	There has been a recent interest on practical aspects and modeling of RIS-assisted communication systems due to their rich use-cases and applications. While the initial studies of \cite{Basar_Access_2019} and \cite{Basar_2019_LIS} and provide useful insights regarding the performance limits of RIS-assisted systems, they do not include path loss effects and consider relatively large RISs with specular reflection only. The subsequent studies of \cite{Ozdogan_2020} and \cite{Garcia_2019} discuss the scattering nature of RIS elements and introduce practical power scaling laws using the principles of physical optics and scattered fields for array near/far-field regions. However, a unified view is presented in \cite{Ellingson} by considering plate scattering and radar range paradigms along with the physical area and practical gain of RIS elements. This study also verifies the fundamental findings of \cite{Tang_2019}, which presents experimental results  on the received signal power involving RISs. Nevertheless, the studies above consider pure line-of-sight (LOS) links between communication terminals and the RIS, limiting their validity in real-world systems. 
	
	More recently, the researchers put forward RIS-assisted mmWave communication systems within the perspective of future networks and reported promising results \cite{Schober_2020,Hey_2020,Yang_2020,Najafi_physics,ntontin_mmWave}. Nevertheless, although  including physical mmWave channel models and massive MIMO architectures, \cite{Schober_2020} does not consider the effect of the RIS on the mmWave channel and models the RIS as an access point with a feed antenna. Similarly, \cite{Hey_2020} assumes an RIS in the form of a uniform linear array (ULA), which might be difficult to implement in practice. Moreover, the authors  deals with point-to-point mmWave links only with ULA-type RISs in \cite{Yang_2020}.  In \cite{Najafi_physics}, a scalable optimization framework is presented by considering a large RIS based on a physics-based model.
Finally, the optimal placement for the RIS-assisted transmission is discussed in mmWave point-to-point link in \cite{ntontin_mmWave}.
	
	Against this background, we observe that there is an urgent need for a physical, open-source, and widely applicable mmWave channel model to be used in various RIS-assisted systems in indoor and outdoor environments. Considering that RIS channel modeling is the first step towards standardized RIS-empowered networks, this paper aims to create a new line of research by integrating RISs into state-of-the-art 5G physical channel models and provide a solid baseline for practical implementation campaigns with sophisticated RIS designs under the far-field assumption. This paper also paves the way for more sophisticated MIMO and time-varying models. Within this context, our major contributions are summarized below:
	\begin{itemize}
		\item We introduce a fundamental channel model by using power scaling laws for RIS-assisted systems in the presence of multiple scatterers and formulate a baseline cascaded physical channel model under the far-field assumption.
		\item Considering the 5G mmWave channel model with random number of clusters/scatterers and the characteristics of the RIS, we provide a unified narrowband channel model for RIS-assisted systems in indoor and outdoor environments for the first time. This model includes many physical characteristics such as LOS probability, shadowing effects, and shared clusters. More importantly, our model considers realistic gains and array responses for RIS elements in addition to the existing channel models.
		\item Using our comprehensive channel model, we demonstrate the potential use-cases and promising gains of RIS-assisted communication in certain setups and provide useful guidelines towards the effective use of RISs.
		\item We introduce the open-source \textit{SimRIS Channel Simulator} MATLAB package, which can be used in channel modeling of RIS-based systems with tunable operating frequency, terminal locations, number of RIS elements, and environments. 
	\end{itemize}
	
The rest of the paper is organized as follows. In Section II, we present the system model of RIS-assisted communication and introduce our baseline channel model. In Sections III and IV, we put forward our RIS channel model for indoors and outdoors, respectively. Section V summarizes the major steps of RIS-assisted channel modeling, while Section VI deals with practical RIS architectures, imperfections and spatial correlation for produced channels. Finally, our numerical results are presented  in Section VII, and the paper is concluded in Section VIII.

\section{RIS-Assisted Communications: System Model}
In this section, we present a simplified and deterministic system model by considering the link power equation and the channel model of an RIS-assisted system with obstacles or reflecting/scattering elements (interacting objects, IOs) between the terminals and the RIS. It is assumed that there are $M$ IOs (scatterers) between the transmitter (Tx) and the RIS as shown in Fig. 1, while assuming a pure LOS link between the RIS and the receiver (Rx). We also assume unit transmit power and antenna gains for clarity of presentation. Here, $a_m$, $b_{m,n}$, and $c_n$ respectively stand for the distances between Tx and the $m$th IO, the $m$th IO and the $n$th RIS element, and the $n$th RIS element and Rx. Furthermore, radar cross section (RCS, in $\text{m}^2$) of the $m$th IO is shown by $\sigma_{\text{RCS}}^m$, and the gain of the corresponding RIS element is assumed to be $G_e$, while a generalized element radiation pattern is considered in the sequel.

\begin{figure}[!t]
	\begin{center}
		\includegraphics[width=0.9\columnwidth]{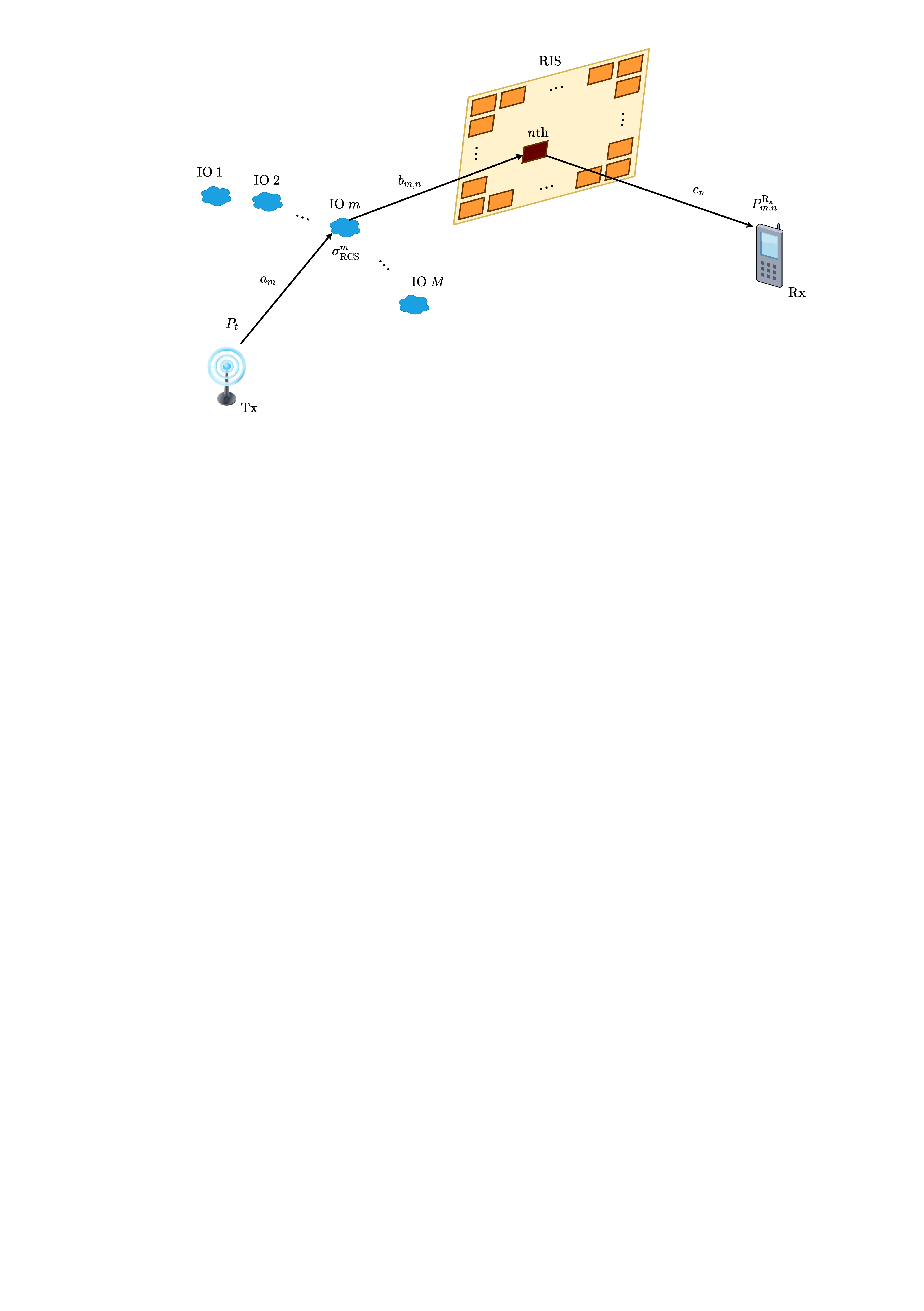}
		\vspace*{-0.2cm}\caption{RIS-assisted communication with $M$ IOs between Tx-RIS.}
		\label{fig:Fig2}
	\end{center} \vspace*{-0.5cm}
\end{figure}

Considering the scattering concept for an RIS with $N$ elements and the direct signal component between Tx and Rx (with or without IOs), the received discrete-time baseband (noise-free) signal can be expressed in vector form as 
\begin{equation}\label{eq:sys_mod}
y= \left( \mathbf{g}^{\mathrm{T}} \mathbf{\Theta} \mathbf{h} + h_{\text{SISO}} \right) x 
\end{equation}
where $\mathbf{g} = \begin{bmatrix}
\sqrt{L_{1}^{\text{LOS}} }  e^{-jkc_1 }& \cdots & \sqrt{L_N^{\text{LOS}} } e^{-jkc_N }
\end{bmatrix}^{\mathrm{T}}$ is the vector of LOS channel coefficients between the RIS and the Rx, $ \mathbf{\Theta} = \mathrm{diag} ( \begin{bmatrix}
\alpha_1 e^{j\phi_1} & \cdots & \alpha_N e^{j\phi_N} 
\end{bmatrix}   ) $ is the matrix of RIS element responses, $\mathbf{h}=\begin{bmatrix}
\sum\limits_{m=1}^{M}\sqrt{L_{m,1}^{\text{RIS}}} e^{-j k (a_m+b_{m,1})} & \cdots & \sum\limits_{m=1}^{M}\sqrt{L_{m,N}^{\text{RIS}}} e^{-j k (a_m+b_{m,N})}
\end{bmatrix}$
is the vector of channel coefficients for the Tx-RIS link composed of $M$ scatterers,  $h_{\text{SISO}}$ characterizes the direct link (narrowband) channel between Tx and Rx, which is equal to $h_{\text{SISO}}=\sqrt{P_{\text{T-R}}} e^{-jkd_{\text{T-R}}}$ for a LOS dominated link, and $x$ is the transmitted signal. Here, $k=2\pi/\lambda$ is the wave number with $\lambda$ being wavelength,  $\alpha_n$ and $\phi_n$ respectively represent controllable magnitude and phase response of the $n$th RIS element, and $L_n^{\text{LOS}}=G_e\lambda^2  / (4\pi c_n)^2$ and $L_{m,n}^{\text{RIS}}=G_e \lambda^2 \sigma_{\text{RCS}}^m/((4\pi)^3 a_m^2 b_{m,n}^2)$ respectively stand for the path gains of the LOS (RIS-Rx) path and the Tx-IOs-RIS path according to the radar range equation \cite{Stutzman}. $P_{\text{T-R}} = \lambda^2 /(4\pi d_{\text{T-R}} )^2$ stands for the received LOS power with $d_{\text{T-R}}$ being the Tx-Rx distance in the presence of a direct link between Tx and Rx. Therefore, the received power via $n$th RIS element at the receiver can be obtained as $P_{m,n}^{\text{Rx}}=L_n^{\text{LOS}} L_{m,n}^{\text{RIS}}$. It is worth noting that when the RIS is far from the Tx and the Rx, $b_{m,n}$ and $c_n$ may be assumed to be independent of the RIS element $n$ from the perspective of channel gains, but not from the phases. We also note that the overall path gain (attenuation) is  obtained as the product of the path gains of individual paths, which are related by $ 1/a_m^2 $, $ 1/b_{m,n}^2 $, and $ 1/c_n^2 $, respectively.

The signal model of \eqref{eq:sys_mod} can be used to assess the power budget of an RIS-assisted system operating in a more realistic environment with multiple IOs compared to the initial LOS-dominated model of \cite{Ellingson}. However, this model again does not capture the variations of the environmental objects and transmit/receiver movements, i.e., based on a static setup.

\section{Physical Channel Modeling: Indoors}
In this section, we will build on the signal model developed in the previous section and provide a unified signal/channel model for RIS-assisted 6G communication systems operating under mmWave frequencies. Our model is generic and can be applied to different environments (indoor/outdoor) and operating frequencies. Here, we focus on indoor channel modeling while the outdoor case is covered in the next section.


It is obvious that the deterministic signal model of \eqref{eq:sys_mod} might be useful for a static environment, i.e., when all IOs as well as the Tx and the Rx are stationary. On the other hand, for a dynamic environment, we need to resort to statistical channel models. For instance, for large number of randomly distributed IOs, the summations in the entries of the RIS-assisted channel vector go to complex Gaussian distribution, yielding a Rayleigh-type amplitude distribution for the received signal, even with fixed terminals. However, the RIS-based channel model involves more sophisticated fading phenomena (double Rayleigh) when both terminals are moving.

The important question is how the channel amplitudes and phases can be modeled in a real-world setup involving an RIS with adjustable phase shifts. To answer this question, there is a need for the development of a unified statistical channel model to be used in RIS-assisted communications. For this purpose, we revisit and apply the well-known statistical channel models to our general signal model of \eqref{eq:sys_mod}. 

Considering the promising potential of RIS-assisted systems for mmWave communications, we build our framework on the clustered statistical MIMO model, which is widely used in 3GPP standardization \cite{3GPP_5G}, while a generalization is possible. In the following three subsections, we present our solutions to generate Tx-RIS, RIS-Rx and Tx-Rx subchannels.


\subsection{Tx-RIS Channel $(\mathbf{h})$}
\begin{figure}[!t]
	\begin{center}
		\includegraphics[width=1\columnwidth]{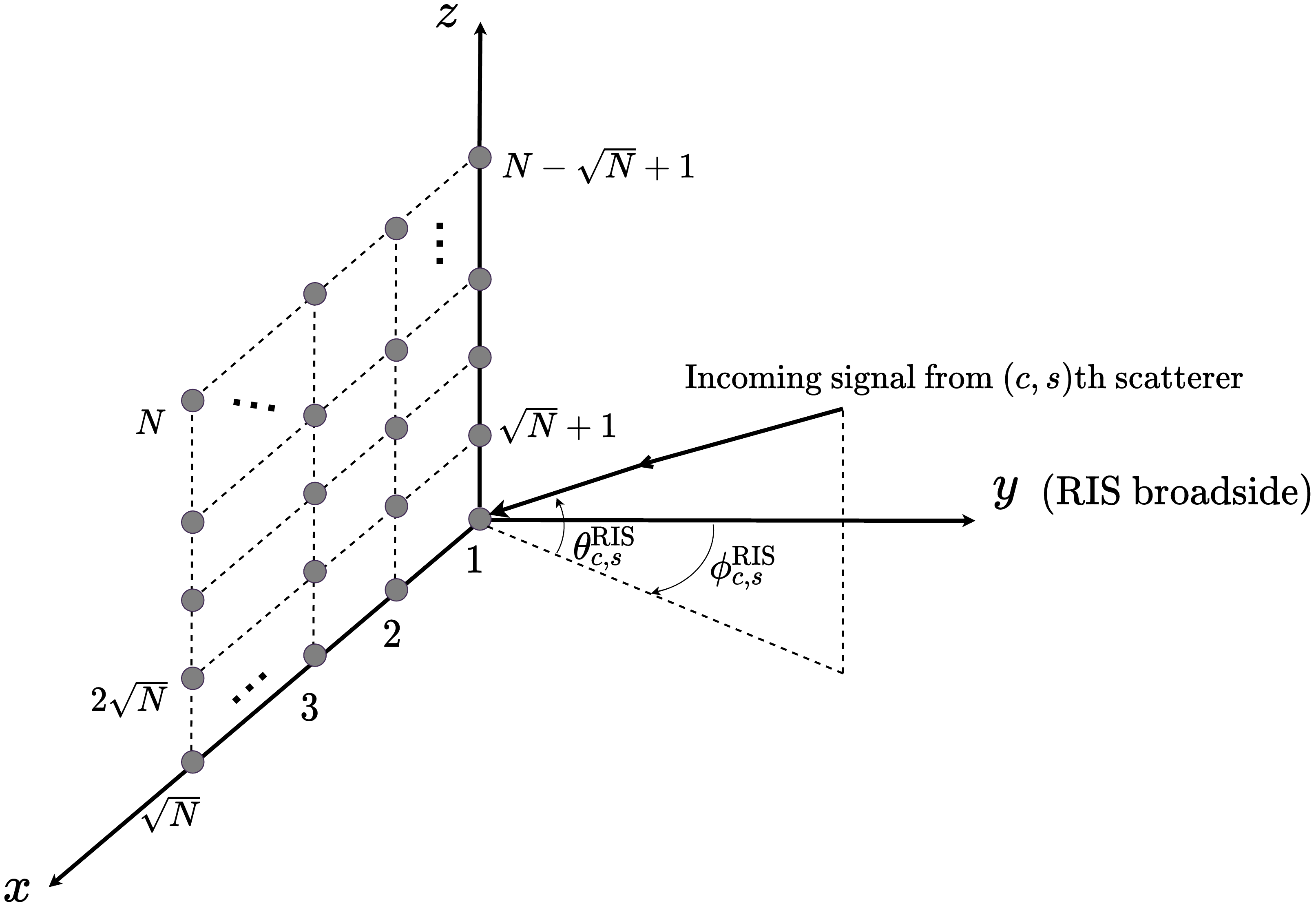}
		\vspace*{-0.6cm}\caption{3D array response geometry for a square RIS with $N$ elements.}
		\label{fig:Fig3}
	\end{center}\vspace*{-0.5cm}
\end{figure}

We assume that the total number of $M$ IOs are grouped under $C$ clusters, each having $S_c$ sub-rays for $c=1,\ldots,C$, that is $M=\sum_{c=1}^{C}S_c$. A cluster can be simply defined as a group of sub-rays that share the same spatial and/or temporal characteristics \cite{Hemadeh_2018}.  In light of this information, the vector of Tx-RIS channel coefficients $\mathbf{h} \in \mathbb{C}^{N\times 1}$ can be rewritten for a clustered model by considering array responses and path attenuations:
\begin{equation}\label{eq:13}
\mathbf{h}=
\gamma \sum\limits_{c=1}^{C} \sum\limits_{s=1}^{S_c} \beta_{c,s} \sqrt{G_e(\theta_{c,s}^{\text{RIS}})L_{\text{T-RIS}}} \,\, \mathbf{a} ( \phi_{c,s}^{\text{RIS}}, \theta_{c,s}^{\text{RIS}} ) + \mathbf{h}_{\text{LOS}}
\end{equation}
where $ \gamma= \sqrt{\frac{1}{\sum\nolimits_{c=1}^{C} S_c}}$ is a normalization factor, commonly used in clustered channel models \cite{3GPP_5G}, $\mathbf{h}_{\text{LOS}}$ is the LOS component to be discussed later, $\beta_{c,s} \sim \mathcal{CN} (0,1)$ and $L_{\text{T-RIS}}$ respectively stand for the complex Gaussian distributed path gain and attenuation associated with the Tx-RIS link, and $G_e(\theta_{c,s}^{\text{RIS}})$ is the rotationally symmetric RIS element pattern \cite{Nayeri} in the direction of the $(c,s)$th scatterer. Here,  $\mathbf{a} ( \phi_{c,s}^{\text{RIS}}, \theta_{c,s}^{\text{RIS}} ) \in \mathbb{C}^{N\times 1}$ is the array response vector\footnote{The same array response is valid when the RIS is placed either on the $xz$ or $yz$ planes since elevation and azimuth angles are defined with respect to the array broadside not to the axes. However, this array response should be modified if the RIS lies on the $xy$ plane or on another surface that is tilted with respect to the global coordinate system.} of the RIS for the considered azimuth and elevation arrival angles (with respect to the RIS broadside) and given as follows for uniformly distributed RIS elements with an inter-element spacing of $d$ \cite{Balanis}:
\begin{align}\label{eq:14}
\mathbf{a} ( \phi_{c,s}^{\text{RIS}}, \theta_{c,s}^{\text{RIS}} ) &= \left[ 
1 \quad \cdots \quad e^{jkd ( x \sin \theta_{c,s}^{\text{RIS}} + y\sin  \phi_{c,s}^{\text{RIS}} \cos \theta_{c,s}^{\text{RIS}}) } \cdots \right. \nonumber \\
&\hspace*{-0.3cm}\left. e^{jkd ( (\sqrt{N}-1) \sin \theta_{c,s}^{\text{RIS}} + (\sqrt{N}-1)\sin  \phi_{c,s}^{\text{RIS}} \cos \theta_{c,s}^{\text{RIS}}) } \right]^{\mathrm{T}}
\end{align}
where $0 \le x \le \sqrt{N}-1$ and $0 \le y \le \sqrt{N}-1$\footnote{For simplicity, we assume a square RIS with $\sqrt{N}$ elements at both horizontal and vertical axes while a generalization is straightforward.}. To obtain the array response in \eqref{eq:14}, we consider the 3D RIS geometry given in Fig. \ref{fig:Fig3}, where the RIS elements are counted from right to left and from bottom to up, and the RIS element at the southeast is taken as the reference, that is, corresponds to the first entry of the array response vector. 

\begin{figure*}[!t]
	\begin{center}
		\includegraphics[width=1.8\columnwidth]{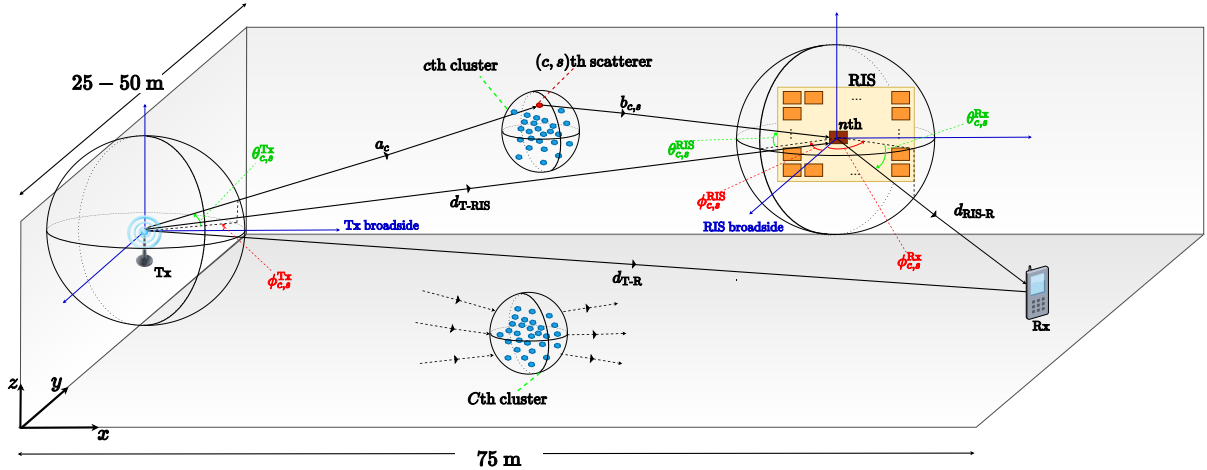}
		\vspace*{-0.3cm}\caption{Generic InH Indoor Office environment with $C$ clusters between Tx-RIS and an RIS mounted in the $xz$ plane (side wall).}
		\label{fig:Fig4}
	\end{center}\vspace*{-0.5cm}
\end{figure*}

To model the RIS element radiation $G_e(\theta_{c,s}^{\text{RIS}})$, we consider the $\cos^q$ pattern, which is widely used for reflectarrays as well as feed patterns \cite{ Nayeri}. In light of this, we consider
\begin{equation}\label{eq:Ge}
G_e(\theta_{c,s}^{\text{RIS}})= 2(2q+1) \cos^{2q}(\theta_{c,s}^{\text{RIS}}), \quad -\pi/2 <\theta_{c,s}^{\text{RIS}}<\pi/2
\end{equation}
where $2(2q+1)$ is a normalization term used for energy conservation, i.e., the integral of $G_e(\theta_{c,s}^{\text{RIS}})$ over a surface enclosing the element is $4\pi$ steradian: 
\begin{equation}\int_{\phi=0}^{2\pi} \int_{\theta=0}^{\pi/2} G_e(\theta) \sin \theta d\theta d\phi=4\pi
\end{equation}
Although being idealistic, this pattern not only is easy to work with but also presents the major part of the main lobe of many realistic antennas \cite{Balanis}. Following the steps of \cite{Ellingson}, we consider $q=0.25 G_e(0)-0.5$, where $G_e(0)$ is calculated from $G_e(0)= 4\pi A_e(0)  / \lambda^2$. Considering the physical area of an RIS element given by $A_e(0)= (\lambda/2)^2$, we obtain $q=0.285$, which corresponds to an element gain of $G_e(0)=\pi$ ($5$ dBi). It is worth noting that although this gain value is consistent with the ones used for patch antennas in practice, the parameter $q$ can be tuned as well by increasing the physical area of RIS elements. We also note that all elements have the same radiation pattern, which is also a reasonable assumption for electrically-large RISs \cite{Ellingson}.

In our model, the number of clusters ($C$), number of sub-rays per cluster ($S_c,c=1,\ldots,C$), and the locations of the clusters can be determined according to the given model and wireless application. As suggested by \cite{Hemadeh_2018} and the references there-in, the number of clusters in mmWaves is typically modeled by Poisson distribution $C \sim \mathcal{P} (\lambda_p)$, whose variance $\lambda_p$ is determined according to a given scenario and operating frequency\footnote{To ensure that at least one cluster exists in the environment, we consider $C = \max \left\lbrace 1,\mathcal{P} (\lambda_p) \right\rbrace $. }. As noted in \cite{Hemadeh_2018,Andrea_2016}, $\lambda_p=1.8$ can be considered for the $28$ GHz band, while $\lambda_p=1.9$ is a suggested value for $73$ GHz. Following \cite{Rappaport_2015}, we assume that the number of sub-rays for the $c$th cluster in an integer uniformly distributed between $1$ and $30$, i.e., $S_c \sim \mathcal{U}[1,30]$, which is a reasonable assumption for both $28$ and $73$ GHz bands. For a given reference cluster $c$, the azimuth departure angles ($\phi^{\text{Tx}}_{c,s},s=1,\ldots,S_c$) of the available sub-rays are assumed to be conditionally Laplacian distributed with a mean value $\phi^{\text{Tx}}_{c}$ following $\mathcal{U}[-\pi/2,\pi/2]$ distribution \cite{Andrea_2016}. Similarly, the elevation departure angles  $(\theta^{\text{Tx}}_{c,s},s=1,\ldots,S_c)$ are conditionally Laplacian with a mean value $\theta^{\text{Tx}}_{c} \sim \mathcal{U}[-\pi/4,\pi/4]$\footnote{This  modification has been made in the interval of the target uniform distribution to have more evenly distributed clusters, particularly for indoor environments. However, more sophisticated distributions can be considered in our model for azimuth/elevation angles of departure.}. The standard deviation (angular spread) of their Laplacian distribution is set to $\sigma_{\theta}=\sigma_{\phi}=5^{o}$ \cite{Heath_2014}. Consequently, we have $\phi^{\text{Tx}}_{c,s}\sim \mathcal{L}(\phi^{\text{Tx}}_{c},5)$ and $\theta^{\text{Tx}}_{c,s}\sim \mathcal{L}(\theta^{\text{Tx}}_{c},5)$. All these departure angles are given with respect to the Tx broadside. It is worth noting that our model is generic and all these small-scale fading parameters can be adjusted according to the given environment.

\textit{The fundamental difficulty in modeling of a wireless channel involving an RIS arises from the fact that once $C$ clusters, as well as their sub-rays, are randomly generated, the azimuth and elevation arrival angles for an RIS cannot be modeled random as in 3GPP LTE and 5G spatial channel models due to random orientation of mobile receivers}. This can be explained by the fact that the orientation of an RIS, which might typically hang on a wall in indoors or facade of a building in outdoors, is deterministic. Furthermore, unless mounted on a car or UAV, an RIS can be also considered a stationary system entity. As a result, the corresponding departure angles -to be used in the calculation of the RIS array response vector in \eqref{eq:14}- should be calculated considering the random positions of the clusters in channel modeling. This necessities a new approach in mmWave channel modeling for RISs, which will be discussed in the sequel.

For simplicity in calculation of link path losses and arrival angles for the RIS, it is assumed that all scatterers in a given cluster $c$ are at the same distance from the Tx, shown by $a_c$ for $c=1,2,\ldots,C$. Denoting the length of the Tx-RIS LOS link\footnote{This does not necessarily mean that there exists a LOS transmission path between the Tx and the RIS.} 
by $ d_{\text{T-RIS}} $, we assume that $a_c \sim \mathcal{U}[1,d_{\text{T-RIS}}]$, however, for clusters whose angle of departure points toward the ground/ceiling or that fall beyond the side walls (in indoor environments), $a_c$ is reduced. Furthermore, the scatterers that fall outside the walls for indoors or underground for outdoors are ignored. The distances\footnote{Without loss of generality, the first RIS element can be taken as a reference under the far case, i.e., $d_{\text{T-RIS}} > N \lambda/2$ for a square RIS. It is worth noting that the RIS is assumed to be lying in the far field of both Tx and Rx terminals.} between the scatterers and the RIS are shown by $b_{c,s}$ for $c=1,2,\ldots,C$ and $s=1,2,\ldots,S_c$. For given Tx and RIS coordinates $(x^{\text{Tx}},y^{\text{Tx}},z^{\text{Tx}})$ and $(x^{\text{RIS}},y^{\text{RIS}},z^{\text{RIS}})$, we have $d_{\text{T-RIS}}=( (x^{\text{RIS}}-x^{\text{Tx}})^2 + (y^{\text{RIS}}-y^{\text{Tx}})^2 + (z^{\text{RIS}}-z^{\text{Tx}})^2)^{1/2}$. Particularly, let us consider that the Tx lies on the $yz$ plane, while the RIS lies either on the $xz$ plane (Scenario 1 - side wall) or $yz$ plane (Scenario 2 - opposite wall). The coordinates of the $s$th scatterer in $c$th cluster are calculated by $x^{c,s}=x^{\text{Tx}} + a_c\cos\theta^{\text{Tx}}_{c,s}\cos\phi^{\text{Tx}}_{c,s}$, $y^{c,s}=y^{\text{Tx}} - a_c\cos\theta^{\text{Tx}}_{c,s}\sin\phi^{\text{Tx}}_{c,s}$, and $z^{c,s}=z^{\text{Tx}} + a_c\sin\theta^{\text{Tx}}_{c,s}$ for $c=1,\ldots,C$ and $s=1,\ldots,S_c$. From the given coordinates of the scatterers, the distance between each scatterer and the RIS can be easily obtained by  $b_{c,s} =( (x^{\text{RIS}}-x^{\text{c,s}})^2 + (y^{\text{RIS}}-y^{\text{c,s}})^2 + (z^{\text{RIS}}-z^{\text{c,s}})^2)^{1/2}$, while the corresponding RIS arrival angles are obtained by $\phi^{\text{RIS}}_{c,s}=I_{\phi}\tan^{-1} \frac{\left| x^{\text{RIS}}-x^{c,s}\right| }{\left| y^{\text{RIS}}-y^{c,s}\right| } $ and $\theta^{\text{RIS}}_{c,s} = I_{\theta}\sin^{-1} \frac{\left| z^{\text{RIS}}-z^{c,s}\right| }{b_{c,s}}$, where $I_{\phi}=\mathrm{sgn}(x^{\text{RIS}}-x^{c,s})$ and $I_{\theta}=\mathrm{sgn}( z^{c,s}-z^{\text{RIS}})$ for Scenario 1. For Scenario 2, we have $\phi^{\text{RIS}}_{c,s}=I_{\phi}\tan^{-1} \frac{\left| y^{\text{RIS}}-y^{c,s}\right| }{\left| x^{\text{RIS}}-x^{c,s}\right| } $ with $I_{\phi}=\mathrm{sgn}(y^{c,s}-y^{\text{RIS}})$, while the same $\theta^{\text{RIS}}_{c,s}$ of Scenario 1 is valid. Finally, the obtained azimuth and elevation arrival angles ($\phi^{\text{RIS}}_{c,s} $ and $\theta^{\text{RIS}}_{c,s}$) are used in the calculation of the array response vector of \eqref{eq:14}. The considered 3D geometry is given in Fig. \ref{fig:Fig4} as a reference for Scenario 1.

For the attenuation of the corresponding path, we adopt the 5G path loss model (the
close-in free space reference distance model with frequency-dependent path loss exponent, in dB), which is applicable to various environments including Urban Microcellular (UMi) and Indoor Hotspot (InH) \cite{5G_Channel}:
\begin{align}\label{eq:path}
&L_{\text{T-RIS}} = -20\log_{10}\left(\frac{4\pi}{\lambda} \right) \nonumber\\ &\hspace*{0.6cm}-10n\left(1+b\left( \frac{f-f_0}{f_0}\right)  \right) \log_{10}(d_{\text{T-RIS}}) -X_{\sigma}.
\end{align}
Here, $d_{\text{T-RIS}}$ is the distance between the Tx and the RIS, $n$ is the path loss exponent, $b$ is a system parameter, and $f_0$ is a fixed reference frequency (the centroid of all frequencies represented by the path loss model), and $X_{\sigma}\sim \mathcal{N}(0,\sigma^2)$ is the shadow fading term (also known as the shadow factor \cite{Rappaport_2015_b}) in logarithmic units.  We consider the channel parameters reported in \cite{5G_Channel} for InH Office-NLOS/LOS and UMi Street Canyon-NLOS/LOS, which are summarized in Table 1.

\begin{table}[!t]
	\caption{Path Loss Parameters \cite{5G_Channel}}
	\centering
	\begin{tabular}{ll}
		\textit{Scenario}                 & \textit{Parameters }                                                                                             \\\hline
		InH Indoor Office (NLOS) & \begin{tabular}[c]{@{}l@{}}$ n=3.19 $, $ \sigma=8.29 $ dB, \\ $ b=0.06 $, $ f_0=24.2$ GHz \end{tabular}  \\\hline
		InH Indoor Office (LOS)  & \begin{tabular}[c]{@{}l@{}}$ n=1.73 $, $ \sigma=3.02 $ dB,\\ $ b=0 $\end{tabular}                     \\\hline
		UMi Street Canyon (NLOS) & \begin{tabular}[c]{@{}l@{}}$ n=3.19 $, $ \sigma=8.2 $ dB, \\ $ b=0 $\end{tabular}                    \\\hline
		UMi Street Canyon (LOS)  & \begin{tabular}[c]{@{}l@{}}$ n=1.98 $, $ \sigma=3.1 $ dB, \\ $ b=0  $ 	\end{tabular}   \\\hline                 
	\end{tabular}
\end{table}

Finally, the obtained array response vector and link attenuation for each sub-ray is substituted in \eqref{eq:13} to generate NLOS components of $\mathbf{h}$. On the other hand, its LOS component is calculated by
\begin{equation}\label{eq:hLOS}
\mathbf{h}_{\text{LOS}}= I_{\mathbf{h}}(d_{\text{T-RIS}}) \sqrt{G_e(\theta_{\text{LOS}}^{\text{RIS}}) L_{\text{T-RIS}}}  e^{j\eta} \mathbf{a}(\phi_{\text{LOS}}^{\text{RIS}},\theta_{\text{LOS}}^{\text{RIS}})
\end{equation}
where $G_e(\theta_{\text{LOS}}^{\text{RIS}})$ is the RIS element gain in the LOS direction, $\mathbf{a}(\phi_{\text{LOS}}^{\text{RIS}},\theta_{\text{LOS}}^{\text{RIS}})$ is the array response of the RIS in the direction of the Tx, and $\eta \sim \mathcal{U} [0,2\pi]$ is the random phase term. Here $I_{\mathbf{h}}(d_{\text{T-RIS}})$ is a Bernoulli random variable taking values from the set $\left\lbrace 0,1 \right\rbrace $, characterizes the existence of a LOS link for a Tx-RIS separation of $d_{\text{T-RIS}}$. Denoting the frequency independent LOS probability by $p$, i.e., $P(I_{\mathbf{h}}=1)=p$, we can resort to the 5G channel model \cite{5G_Channel} to obtain
\begin{equation}\label{eq:LOS_prob}
p=\begin{cases}
1 & d_{\text{T-RIS}} \le 1.2 \\
e^{-\left(\frac{d_{\text{T-RIS}} -1.2}{4.7} \right) } & 1.2<d_{\text{T-RIS}} \le 6.5 \\
0.32 e^{-\left(\frac{d_{\text{T-RIS}} -6.5}{32.6}\right) } & d_{\text{T-RIS}}>6.5.
\end{cases}
\end{equation}  
It is worth noting that the LOS probabilities of \eqref{eq:LOS_prob} are reported based on intensive measurements in various indoor office environments, most probably, for Rxs at a height of $1$-$1.5$ m and decays very fast with increasing distance. Consequently, for the link between the Tx and the RIS, we resort to \eqref{eq:LOS_prob} if the RIS is located below the level of Tx ($z^{\text{RIS}}<z^{\text{Tx}}$), while we assume $p=1$ for $z^{\text{RIS}}\ge z^{\text{Tx}}$ regardless of $d_{\text{T-RIS}}$. In other words, if the height of the RIS is not smaller than that of the Tx, we assume a clear LOS path between them, which is a reasonable assumption for any choice of $d_{\text{T-RIS}}$. 

\subsection{RIS-Rx Channel $(\mathbf{g})$}
In our indoor communications model, to simply our initial analyses, we assume that the RIS and the Rx are sufficiently close together to have a clear LOS link without noticeable NLOS components in between. According to \eqref{eq:LOS_prob}, a LOS probability of greater than $50$\% is achieved when the inter-terminal distance is less $4.5$ m. We later relax the LOS dominated channel conditions between the RIS and the Rx for outdoor setups. 

For the calculation of LOS-dominated RIS-Rx channel $\mathbf{g}$, we re-calculate the RIS array response from \eqref{eq:14} in the direction of the Rx by calculating azimuth and elevation departure angles $\phi^{\text{RIS}}_{\text{Rx}}$ and $\theta^{\text{RIS}}_{\text{Rx}}$ for the RIS from the coordinates of the RIS and the Rx\footnote{For given RIS and Rx coordinates $(x^{\text{RIS}},y^{\text{RIS}},z^{\text{RIS}})$ and $(x^{\text{Rx}},y^{\text{Rx}},z^{\text{Rx}})$, we have $d_{\text{RIS-R}}=( (x^{\text{Rx}}-x^{\text{RIS}})^2 + (y^{\text{Rx}}-y^{\text{RIS}})^2 + (z^{\text{Rx}}-z^{\text{RIS}})^2)^{1/2}$. For both scenarios, we obtain $\theta^{\text{RIS}}_{\text{Rx}} = I_{\theta}\sin^{-1} \frac{\left| z^{\text{Rx}}-z^{\text{RIS}}\right| }{d_{\text{RIS-R}}}$ for $I_{\theta}=\mathrm{sgn}(z^{\text{Rx}}-z^{\text{RIS}})$. However, for Scenario 1, we have  $\phi^{\text{RIS}}_{\text{Rx}}=I_{\phi}\tan^{-1} \frac{\left| x^{\text{Rx}}-x^{\text{RIS}}\right| }{\left| y^{\text{Rx}}-y^{\text{RIS}}\right| }$ for $I_{\phi}=\mathrm{sgn} (x^{\text{RIS}}-x^{\text{Rx}})$, and for Scenario 2, we have $\phi^{\text{RIS}}_{\text{Rx}}=I_{\phi}\tan^{-1} \frac{\left| y^{\text{Rx}}-y^{\text{RIS}}\right| }{\left| x^{\text{Rx}}-x^{\text{RIS}}\right| }$ for $I_{\phi}=\mathrm{sgn}(y^{\text{Rx}}-y^{\text{RIS}})$.}. The attenuation of these $N$ LOS paths ($L_{\text{RIS-R}}$) is calculated from \eqref{eq:path} by replacing $d_{\text{T-RIS}}$ with $d_{\text{RIS-R}}$ along with other channel related parameters.
The gain of RIS elements is again reflected to $\mathbf{g}$ by considering departure elevation angle $\theta^{\text{RIS}}_{\text{Rx}}$ in \eqref{eq:Ge}. Finally, with $\eta \sim \mathcal{U} [0,2\pi]$,  the vector of LOS channel coefficients can be generated as
\begin{equation}\label{eq:20}
\mathbf{g}=\sqrt{G_e(\theta^{\text{RIS}}_{\text{Rx}}) L_{\text{RIS-R}}} e^{j\eta} \mathbf{a}(\phi^{\text{RIS}}_{\text{Rx}},\theta^{\text{RIS}}_{\text{Rx}}).
\end{equation}

\subsection{Tx-Rx Channel $(h_{\text{SISO}})$}
The RIS-assisted channel has a double-scattering nature, as a result, the single-scattering link between the Tx and Rx has to be taken into account in our channel model. As will be discussed later, even if the RIS is placed near the Rx, the Tx-Rx channel is relatively stronger than the RIS-assisted path, and cannot be ignored in the channel model. 

Since we assume that the RIS and the Rx are relatively closer with a clear LOS path, we may assume that they experience the same clusters. In other words, it is assumed that the distance between the RX and the RIS is smaller than the correlation distance (stationarity interval \cite{5G_Channel}) in indoor environments so that they have common clusters. This also allows us to model the potential correlation  between the channels of RIS and the Rx through shared clusters \cite{Alonzo_2019}.

Using SISO mmWave channel modeling, the channel between these two terminals can be easily obtained (by ignoring arrival and departure angles) as
\begin{equation}\label{eq:18}
h_{\text{SISO}}=
\gamma \sum\limits_{c=1}^{C} \sum\limits_{s=1}^{S_c} \beta_{c,s} e^{j\eta_e} \sqrt{L_{\text{SISO}}} + h_{\text{LOS}}
\end{equation}
where $\gamma$, $C$, $S_c$, and $\beta_{c,s}$ are as defined in \eqref{eq:13} and remain the same for the Tx-Rx channel under the assumption of shared clusters with the Tx-RIS channel, while $h_{\text{LOS}}$ is the LOS component. Here, $L_{\text{SISO}}$ stands for the path attenuation for the Tx-Rx link and $\eta_e$ is the excess phase caused by different travel distances of Tx-RIS and Tx-Rx links over the same scatterers\footnote{Denoting the distances from the RIS and the Rx to the $(c,s)$th scatterer by $b_{c,s}$ and $\tilde {b}_{c,s}$, respectively, we have $\eta_e=k(b_{c,s}-\tilde {b}_{c,s})$.}. 
 $L_{\text{SISO}}$ is calculated from \eqref{eq:path} by considering the same shadow factors.
Similarly, $ h_{\text{LOS}} $ is generated considering the link attenuation for $d_{\text{T-R}}$ if a LOS exists between Tx and the RIS, i.e., we assume that due to relatively close separation of the RIS and the Rx, they simultaneously experience LOS signals with a probability of $p$ \eqref{eq:LOS_prob}. However, if the RIS is positioned at a higher height to have a LOS link with the Tx, we calculate $p$ independently for $h_{\text{SISO}}$.


\section{Physical Channel Modeling: Outdoors}

The comprehensive mmWave channel modeling method of the previous section can be easily extended to outdoor environments by modifying certain system parameters.  Differences in the positions of the RIS and terminals in an outdoor environment will cause changes in channel parameters, while following the same channel modeling strategy. Particularly, path loss exponent $n$ and shadow fading parameter $\sigma$ in \eqref{eq:path} can be adjusted according to the given outdoor propagation environment. In terms of small-scale fading, only for the clusters whose angle of departure points toward the ground, we may reduce the maximum distance by considering terminal distances in the outdoor environment. It is worth noting that our model assumes the placement of the RIS on $xz$ and $yz$ planes, while certain modifications can be made in the array responses and azimuth/elevation angle calculations for the placement of the RIS on the $xy$ plane (on the ground or the roof) or for other tilted/rotated RIS orientations. Following the same steps of Section III.A, $\mathbf{h}$ can be easily generated.

The major change in contrast to our indoor channel model will be in the channel between the RIS and the Rx, which might be subject to small-scale fading as well in outdoor environments with a random number of unique clusters. The LOS-dominated channel of \eqref{eq:20} may be still useful when the distance between the RIS and the Rx is relatively short with a high LOS probability. However, for the general case, we have
\begin{equation}\label{eq:outdoor}
\mathbf{g}=
\bar{\gamma} \sum\limits_{c=1}^{\bar{C}} \sum\limits_{s=1}^{\bar{S_c}} \bar{\beta}_{c,s} \sqrt{G_e(\theta_{c,s}^{\text{Rx}}) L_{\text{RIS-R}}} \,\, \mathbf{a} ( \phi_{c,s}^{\text{Rx}}, \theta_{c,s}^{\text{Rx}} ) + \mathbf{g}_{\text{LOS}}
\end{equation}
where, similar to \eqref{eq:13}, $\bar{\gamma}$ is a normalization term, $\bar{C}$ and $\bar{S_c}$ stand for number of clusters and sub-rays per cluster for the RIS-Rx link, $\bar{\beta}_{c,s}$ is the complex path gain, $ L_{\text{RIS-R}} $ is the path attenuation, $G_e(\theta_{c,s}^{\text{Rx}})$ is the RIS element radiation pattern in the direction of the $(c,s)$th scatterer, $\mathbf{a} ( \phi_{c,s}^{\text{Rx}}, \theta_{c,s}^{\text{Rx}} )$ is the array response vector of the RIS for the given azimuth and elevation angles, and $\mathbf{g}_{\text{LOS}}$ is the LOS component. Similar steps of the previous section can be followed to generate $\mathbf{g}$ in fading environments\footnote{We assume a reduced interval $([\pi/4,\pi,4])$ for the uniform distribution of the mean angle of departure (in azimuth) to ensure that all scatters are within the field of view of the RIS, i.e., not to have scatterers at the back of the RIS.}. Different from indoor scenarios, the LOS probability is also modified as \cite{5G_Channel}
\begin{equation}\label{eq:LOS_prob2}
p=\min(20/d,1)(1-e^{-d/39}) + e^{-d/39}
\end{equation} 
where  $d\in \left\lbrace d_{\text{T-RIS}},d_{\text{RIS-R}},d_{\text{T-R}} \right\rbrace $. This model is adopted from 3GPP/ITU standards and does not include the receiver height as other urban macrocellular models \cite{Haneda_2016}. Furthermore, this model considers ground-level Rxs, which might not completely hold for RIS-assisted links. Nevertheless, since the Tx-RIS and RIS-Rx link lengths are typically larger compared to indoor environments, it is taken as a reference to model the worst-case scenario, while a higher LOS probability is expected for Tx-RIS and RIS-Rx links.

\begin{figure}[!t]
	\begin{center}
		\includegraphics[width=0.9\columnwidth]{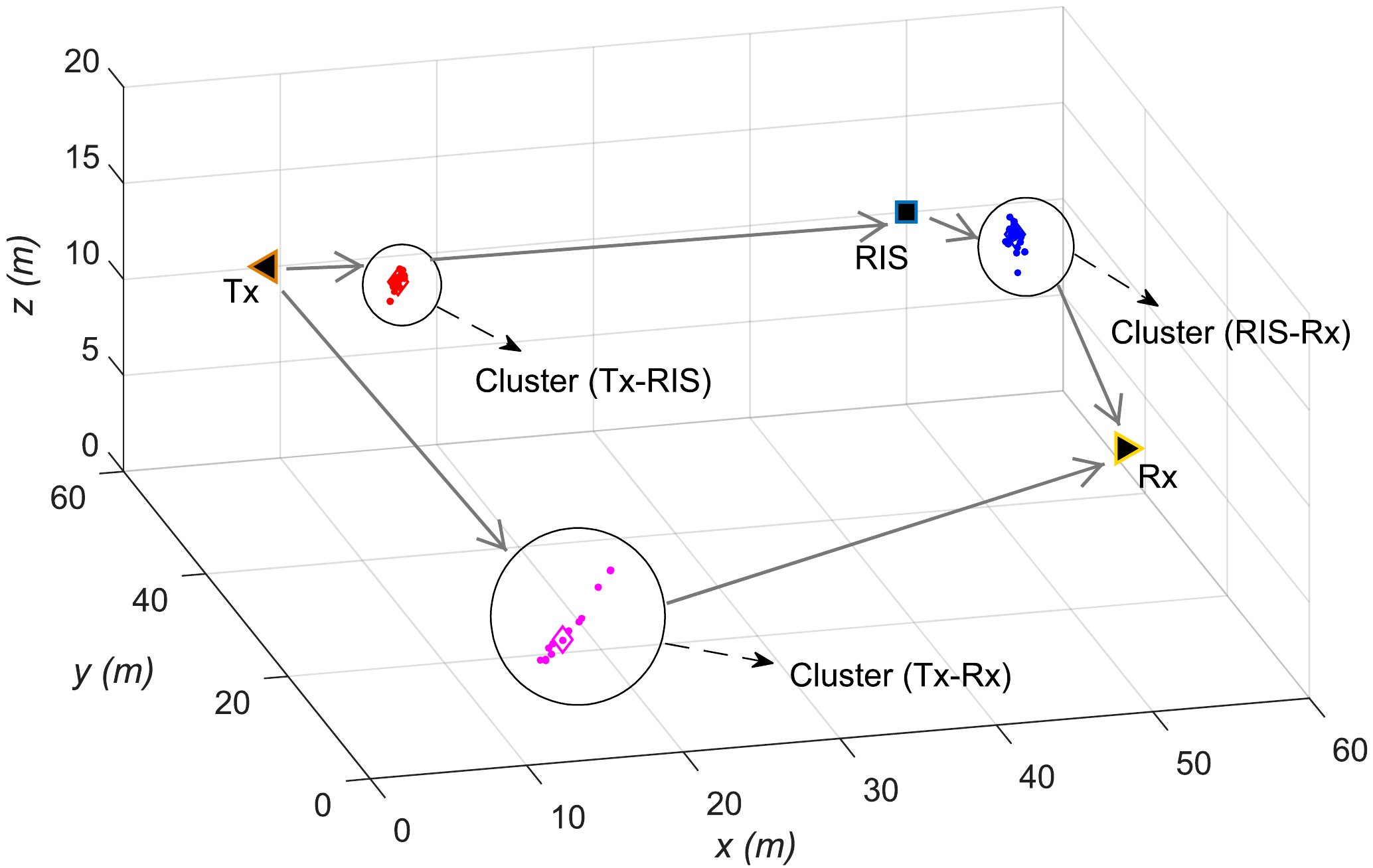}
		\vspace*{-0.3cm}\caption{The considered UMi Street Canyon outdoor environment with random number of clusters/scatterers and an RIS on the $xz$ plane.}
		\label{fig:Out}
	\end{center}\vspace*{-0.5cm}
\end{figure}

For outdoor environments, we assume that the RIS and the Rx are not too close to ensure that they have independent clusters (small scale parameters) as in the 3GPP 3D channel model \cite{3GPP_5G}. In other words, the distance between the Rx and the RIS is not smaller than the correlation distance so that they do not have common clusters. 
Using SISO mmWave channel modeling, the Tx-Rx channel can be easily obtained (by ignoring arrival and departure angles once more) as
\begin{equation}\label{eq:SISO}
h_{\text{SISO}}=
\tilde{\gamma} \sum\limits_{c=1}^{\tilde{C}} \sum\limits_{s=1}^{\tilde{S}_c} \tilde{\beta}_{c,s} \sqrt{L_{\text{SISO}}} + h_{\text{LOS}}
\end{equation}

An example realization of the considered 3D geometry for UMi Street Canyon outdoor environment is given in Fig. \ref{fig:Out} for Scenario 1, where the Tx is mounted at a height of $20$ m, while the Rx is a ground-level user. In this specific 3D geometry, each path has a single cluster with a different number of scatterers, while the number of clusters for each path varies randomly in general.

\section{On RIS-Assisted Channel Modeling with {\normalfont SimRIS}}

In this section, we first summarize the major steps of RIS-assisted channel modeling for indoor and outdoor environments of the previous two sections and then introduce the open-source \textit{SimRIS Channel Simulator}.

The steps of RIS-assisted channel modeling can be summarized as below. Here, steps 1-5 focus on the generation of $\mathbf{h}$, while Steps 6-7 and Step 8 respectively deal with $\mathbf{g}$ and $h_{\text{SISO}}$:

\begin{enumerate}
	\item Give the coordinates of the Tx/Rx and the RIS. Then calculate the direct link distance $d_{\text{T-RIS}}$ between the Tx and the RIS. Calculate the LOS probability for this link and generate the LOS component of $\mathbf{h}$ accordingly using $\phi_{\text{LOS}}^{\text{Tx}}$ and $\theta_{\text{LOS}}^{\text{Tx}}$.
	\item Determine the number of clusters $C$ and sub-rays $S_c$ for $c=1,\ldots,C$. Generate azimuth and elevation departure angles $\phi_{c,s}^{\text{Tx}}$ and $\theta_{c,s}^{\text{Tx}}$, and the distances between the Tx and the clusters.
	\item Calculate the distances between scatterers and the RIS $b_{c,s}$.  Calculate the angles of arrival $\phi^{\text{RIS}}_{c,s}$ and $\theta^{\text{RIS}}_{c,s}$ for the RIS with respect to the RIS broadside.
	\item Calculate the array response vector $ \mathbf{a} ( \phi_{c,s}^{\text{RIS}}, \theta_{c,s}^{\text{RIS}} )$ for all $c$ and $s$.
	\item Calculate the link attenuation $L_{\text{T-RIS}}$ for given system parameters and complex path gains $\beta_{c,s}$. Generate the vector of channel coefficients $\mathbf{h}$ for the Tx-RIS link.
	\item Calculate the LOS distance $d_{\text{RIS-R}}$ as well as LOS azimuth and elevation departure angles  $\phi^{\text{RIS}}_{\text{Rx}}$ and $\theta^{\text{RIS}}_{\text{Rx}}$ for the RIS. 
	\item For indoor environments, re-calculate the array response vector by $\phi^{\text{RIS}}_{\text{Rx}}$ and $\theta^{\text{RIS}}_{\text{Rx}}$, and generate the vector of LOS channel coefficients $\mathbf{g}$ between the RIS and the Rx. For outdoor environments, determine the number of clusters $\bar{C}$ and sub-rays $\bar{S_c}$ and repeat corresponding procedures under Steps 2-5 (without calculating arrival angles and array response (Step 4) at the Rx due to its single antenna) for the RIS-Rx channel by also considering its LOS component.
	\item Generate the direct link channel coefficient $h_{\text{SISO}}$ by considering the same clusters with the Tx-RIS link for indoors and independent clusters for outdoors. For indoors, calculate the excess phase $\eta_e$ and link distances only to obtain $h_{\text{SISO}}$. For outdoors, determine the number of clusters $\tilde{C}$ and sub-rays $\tilde{S_c}$ and generate $h_{\text{SISO}}$. Consider its LOS component for all cases.
	\item Considering the matrix of RIS responses $\mathbf{\Theta}$ (to be discussed in Section VI), generate the overall channel coefficient $h=\mathbf{g}^{\mathrm{T}} \mathbf{\Theta} \mathbf{h} + h_{\text{SISO}}$.
\end{enumerate}
Following the main steps given above, for InH Indoor Office and UMi Street Canyon environments, $ \mathbf{h}$, $\mathbf{g}$ and $h_{\text{SISO}}$ channels can be produced by performing Monte Carlo simulations at $28$ and $73$ GHz frequencies in \textit{SimRIS Channel Simulator} \cite{SimRIS_latincom}. Number of RIS elements ($N$) and number of channel realizations can be defined as user-selectable parameters as well as the Tx, Rx and RIS locations by considering our 3D geometry. Furthermore, the RIS position can be selected for $xz$ plane (side wall) or $yz$ plane (opposite wall) for both environments. \textit{SimRIS Channel Simulator} MATLAB package is provided as a companion of this paper, and interested readers are referred to \cite{SimRIS_latincom} for the details of channel generation and the use of this simulator.

\section{Practical Issues and Channel Correlation}
In this section, we discuss the effects of different operating modes of an RIS on the system performance and importance of effective RIS positioning. Additionally, we analyze the correlation of the produced Tx-RIS channel via SimRIS Channel Simulator, while a generalization to the RIS-Rx channel is straightforward.

The fundamental benefit of an RIS, in other words, ``intelligence'' of an RIS stems from the software-controlled phase adjustment of its reflecting elements by allowing the optimization of desired system performance metrics (e.g., energy efficiency, transmit power, and achievable rate). Here, while it is assumed that the RISs is in the form of a software-controlled uniform planar array as in Fig. \ref{fig:Fig3}, it is possible to generate channel models for different RIS designs by following the steps of RIS-assisted channel modeling.

By considering the received signal of \eqref{eq:sys_mod} in a noisy environment, the received instantaneous signal-to-noise ratio (SNR) is given by
\begin{equation}\label{eq:rSNR}
\rho=\frac{|\mathbf{g}^T\mathbf{\Theta}\mathbf{h}+ h_{\text{SISO}}|^2P_t}{P_N}
\end{equation}
where $P_N$ denotes the noise power. If perfect phase knowledge is considered at the RIS, \eqref{eq:rSNR} can be maximized by adjusting the phases of the RIS-assisted path to the phase of the direct path between Tx and Rx \cite{Basar_2019_LIS}.

Imperfections in phase knowledge acquirement should be taken into consideration due to hardware limitations encountered in real-world applications. Two major types of imperfections that may occur in real-time applications are phase estimation errors and quantization errors.  Zero-mean von Mises distribution can be used to model the phase estimation errors with the concentration parameter $ \kappa $, which is a metric for the estimation accuracy \cite{Coon_2019}. Quantization errors result from a limited number of discrete phase shifts of RIS elements in the absence of high-resolution phase shifters with infinite-level. If a finite-level phase shifter is considered, $q$ control bits can be used to adjust $2^q$ discrete phases. These discrete phases will cause the quantization errors, which follow uniform distribution over $[-2^{-q} \pi,2^{-q} \pi]$. Although it is not possible to construct  $ \mathbf{\Theta} $ that will fully align the channel phases under these system imperfections, it will still be possible to achieve performance improvements with optimal and near-optimal phase adjustments \cite{Abeywickrama_2019}.

Although it is not possible to construct $ \mathbf{\Theta} $ that will optimize \eqref{eq:rSNR} under these system imperfections, it will still be possible to achieve performance improvements with near-optimal phase adjustments. 

As will be discussed in the next section, in order to exploit RISs to boost the communication system performance, they have to positioned very carefully. Considering the LOS probability between the Tx and RIS, a reliable transmission can be provided by placing the RIS to obtain a LOS link for the Tx-RIS channel. In order to boost the performance of existing communication systems, it is necessary to place RISs on the walls closest to Rxs in order to ensure that the Tx and RIS have also an LOS link. For both indoors and outdoors, positioning the RIS on the  $xz$ plane (side wall) or $yz$ plane (opposite wall) is the most reasonable solution to meet these criteria.  

\begin{figure} 
	\includegraphics[width=1\columnwidth]{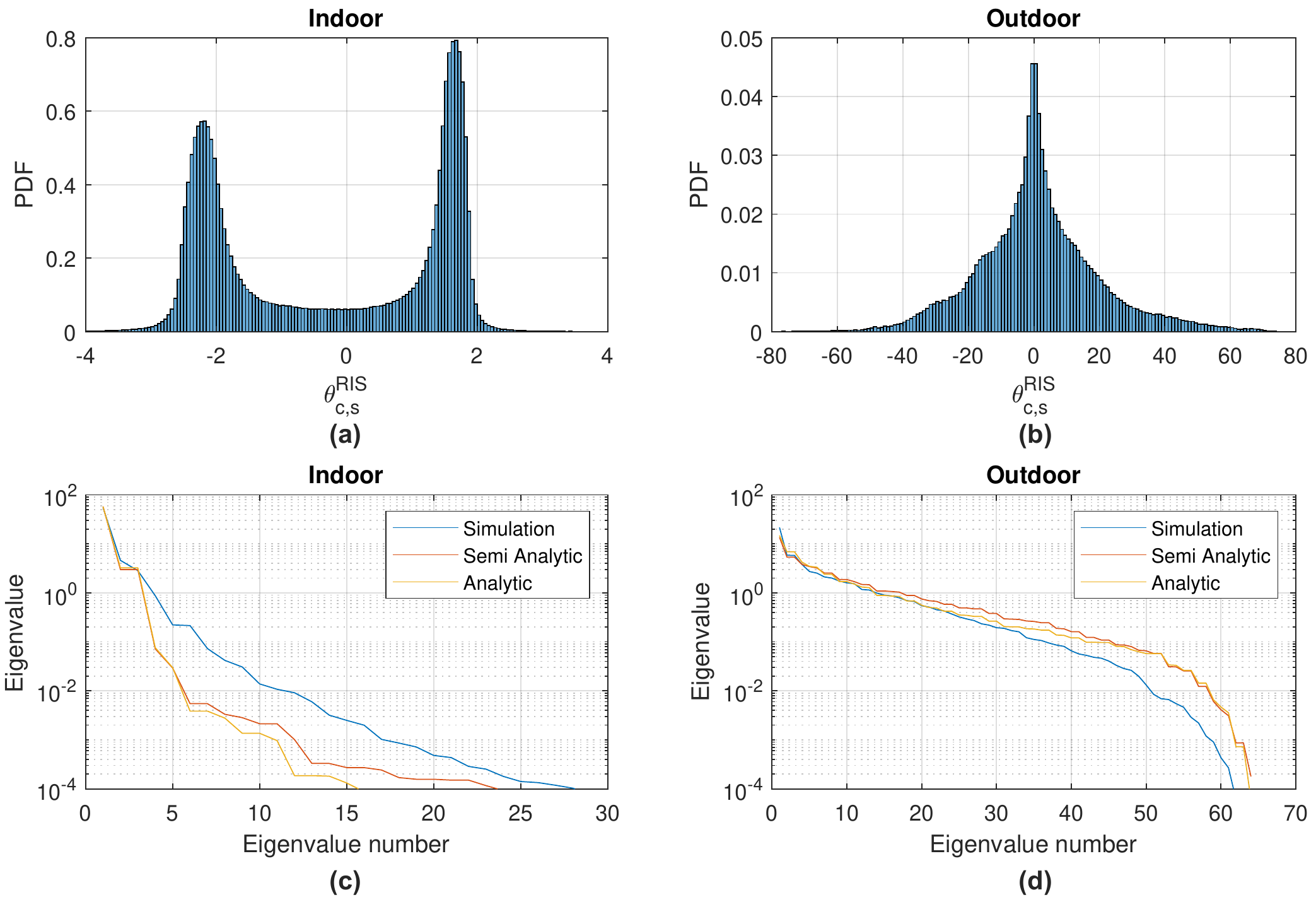} 
	\caption{The PDF of $(\theta_{c,s}^\text{RIS})$ in an (a) indoor and (b) outdoor environment. Eigenvalue spread of spatial correlation matrix in an (c) indoor and (d) outdoor environment.}
	\label{Fig:Cor} \vspace*{-0.4cm}
\end{figure}

Additionally, we analyze the spatial correlation between two arbitrary RIS elements by ignoring the LOS component of the Tx-RIS channel given in \eqref{eq:13}. We assume that all $\theta_{c,s}^\text{RIS}$ and $\phi_{c,s}^\text{RIS}$ are approximately the same for scatterers within the same cluster. Therefore, the elevation and azimuth angles are respectively simplified as $\theta_{c}^\text{RIS}$ and $\phi_{c}^\text{RIS}$, where $ c $ stands for the index of the corresponding cluster. In order to perform a spatial correlation analysis, since $G_e(\theta_{c,s}^\text{RIS})$ and $L_\text{T-RIS}$ have relatively small variations, we use the expected value of the term in square root in \eqref{eq:13} by considering all scatterers in the environment and define it as  $\mu=\mathrm{E}\left[\sqrt{G_e(\theta_{c,s}^\text{RIS})L_\text{T-RIS}}\right]$. A finite number of multipath components is obtained  due to the limited number of clusters in the transmission environment, and this leads to a non-isotropic scattering environment.  The channel coefficient for the $n$th element of the RIS can be approximated by
\begin{equation}\label{eq:coef_corr}
{h}_{n}\approx
\gamma\mu\sum\limits_{c=1}^{C} \sum\limits_{s=1}^{S_c} \beta_{c,s} e^{jkd(n_z\sin(\theta_c^\text{RIS})+n_x\sin(\phi_c^\text{RIS})\cos(\theta_c^\text{RIS}))}
\end{equation}
where the coordinates of this element are respectively denoted by $(n_x,n_z)$ in horizontal and vertical axes of the RIS by considering the RIS geometry in Fig. 2. 
The  correlation matrix of $\mathbf{h}$ is represented by $\mathbf{R}$, therefore, the correlation coefficient of $n$th and $m$th element of the RIS is calculated by
\begingroup\makeatletter\def\f@size{8.5}\check@mathfonts\
\begin{align} \label{eq:Cor}
&\mathbf{R}_{n,m}=\frac{1}{A}\mathrm{E}[ {h}_{n}{h}_{m}^*]  =\frac{1}{A}\mathrm{E}\Big[\sum\limits_{i=1}^{C}\sum\limits_{j=1}^{C}\bar{\mu}_i\bar{\mu}_j^*\nonumber \\ &\times e^{jkd[n_z\sin(\theta_i^\text{RIS})-m_z\sin(\theta_j^\text{RIS})
+n_x\sin(\phi_i^\text{RIS})\cos(\theta_i^\text{RIS})-m_x\sin(\phi_j^\text{RIS})\cos(\theta_j^\text{RIS})]}\Big]
\end{align}\endgroup 
where $A$ denotes the normalization coefficient and $\bar{\mu}_c=\gamma\, \mu(\beta_{c,1}+\dots+\beta_{c,S_c})$ for $c\in\left\lbrace 1,\dots,C\right\rbrace$. Since $\beta_{c,s}$ are i.i.d. normal random variables, $\mathrm{E}[\bar{\mu}_i\bar{\mu}_j^*]=0$ for $i\neq j$. Considering that $n$th and $m$th elements of the RIS are located in the same row and $\text{E}[\left|\bar{\mu}_c\right| ^2 ]=\gamma^2\mu^2S_c $ for $c\in\left\lbrace 1,\dots,C\right\rbrace$, \eqref{eq:Cor} is simplified as
\begin{equation}\label{eq:Cor2}
\mathbf{R}_{n,m}=
\frac{1}{A}\gamma^2\mu^2 \sum\limits_{c=1}^{C} S_c \mathrm{E}\left[e^{jkd_{n,m}\sin(\theta_c^\text{RIS})}\right] 
\end{equation}
where $d_{n,m}$ represents the distance between $n$th and $m$th reflecting elements. If these reflecting elements are not located in the same row, it can be assumed that the coordinate system of the RIS is rotated such that $n$th and $m$th reflecting elements are aligned in the same row and the distance between them becomes the horizontal distance as in \cite{Emil_hardening}. Since it is hard to obtain the individual $ \theta_c^\text{RIS} $ distribution  for each cluster, $\theta_c^\text{RIS}$ should be modeled as the same for all $ c $. Since $\gamma$ is also a normalization term based on the number of scatterers as expressed in (2), we assume the normalization constant $A=\mu^2$, so that $\mathbf{R}_{n, m}=1 $ for $n=m$. Thus, considering the same  distribution for all  $ \theta_c^\text{RIS} $ in (16), the correlation coefficient between two reflecting elements is represented as follows
\begin{align}\label{eq:Cor6}
\mathbf{R}_{n,m}=\mathrm{E}[e^{jkd_{n,m}\sin(\theta_c^\text{RIS})}].
\end{align}
Here, the distribution of $\theta_{c}^\text{RIS}$ depends on the Tx and RIS locations, and the constraints in considered transmission environment. Therefore, it is reasonable to obtain a distribution by considering a large number of $\theta_{c}^\text{RIS}$ samples for fixed positions of Tx and RIS. Since it is difficult to express the distribution of $\theta_c^\text{RIS}$ mathematically regardless of the Tx and the RIS positions, the correlation coefficient between $n$th and $m$th elements can be obtained semi-analytically by taking the mean of many samples from the distribution of $\theta_c^\text{RIS}$ as given below.
\begin{align}\label{eq:Cor3}
\mathbf{R}_{n,m}^s=\frac{1}{N_\theta}\sum_{i=1}^{N_\theta} e^{jkd_{n,m}\sin(\theta_i^\text{RIS})}
\end{align}
where $N_\theta$ represents the number of samples and $\theta_i^\text{RIS}$ is $i$th sample from the distribution of $\theta_c^\text{RIS}$.

To get further insights, we obtain the probability density functions (PDFs) of $\theta_{c,s}^\text{RIS}$ for indoor and outdoor environments by performing Monte Carlo simulations via SimRIS Channel Simulator as shown in Figs. \ref{Fig:Cor}(a) and (b) for $N=64$. Here, the Tx and the RIS are respectively located at $ (0,25,2) $ and $ (40,50,2) $ for an indoor environment, while $ (0,25,40) $ and $ (70,85,40) $ for an outdoor environment. Since we assume that $\theta_{c,s}^\text{RIS}$ is the same for all scatterers within the same cluster, the distribution of $\theta_c^\text{RIS}$ is assumed as the same with $\theta_{c,s}^\text{RIS}$. 
Based on this assumption, we can approximately express the PDFs of $\theta_c^\text{RIS}$ for outdoor and indoor environments, respectively, as follows\footnote{The PDFs in \eqref{eq:pdf1} and \eqref{eq:pdf2} are only used for modeling assumed distribution of   $\theta_c^\text{RIS}$ according to Figs. \ref{Fig:Cor}(a) and \ref{Fig:Cor}(b).}:
\begin{align}\label{eq:pdf1}
f_{\theta}(\theta_c^\text{RIS})=\frac{1}{22}e^{\frac{-|\theta_c^\text{RIS}|}{11}}, & -\Delta\theta \leq \theta_c^\text{RIS} \leq \Delta\theta,
\end{align}
\begin{align}\label{eq:pdf2}
f_{\theta}(\theta_c^\text{RIS})=
\begin{cases} 
\frac{2.5}{1.7/0.7}e^{-(\theta_c^\text{RIS}+2.3)2.5\,\xi(0.7)^\xi}, & 0<\theta_c^\text{RIS}\leq \Delta\theta \\
\frac{3.5}{2.4/1.4}e^{-(\theta_c^\text{RIS}-1.7)3.5\xi(1.4)^\xi}, & \hspace*{-0.25cm}-\Delta\theta \leq \theta_c^\text{RIS} \leq 0 \\
\end{cases}.
\end{align}
In \eqref{eq:pdf1}$, \theta_c^\text{RIS}$ follows a Laplacian distribution with zero-mean and $11^{o}$ standard deviation (angular spread) as $\theta_c^\text{RIS}\sim\mathcal{L}(0,11)$ for outdoors, since it depends on $\theta_{c,s}^\text{Tx}$ which is a Laplacian distributed random variable as well. Unlike outdoors, the clusters are relatively restricted between the ground and ceiling of the indoor environment, and accumulated in narrow regions. Therefore, $\theta_{c,s}^\text{RIS}$ has a very narrow range as seen in Fig. \ref{Fig:Cor}(a) due to the geometry of the indoor office. In an indoor environment, the distribution of $\theta_c^\text{RIS}$, with $\Delta\theta$ being the boundary of two different distributions as seen from \eqref{eq:pdf2}, can be modeled by asymmetric Laplacian distribution as $\theta_c^\text{RIS} \sim \mathcal{AL}(\tau,\epsilon,\zeta)$, where $\tau$, $\epsilon$ and $\zeta$ stand for the location, scale and asymmetry parameters, respectively. Moreover, the distribution for indoors is approximated as a pairwise function, $\theta_c^\text{RIS}\sim \mathcal{AL}(-2.3,2.5,0.7)$ for $0<\theta_c^\text{RIS}\leq \Delta\theta $ and $\theta_c^\text{RIS}\sim \mathcal{AL}(1.7,3.5,1.4)$ for $-\Delta\theta \leq \theta_c^\text{RIS} \leq 0$ as in \eqref{eq:pdf2}, where $\xi=\mathrm{sign}(\theta_c^\text{RIS}-\tau)$.
By using the approximated distribution functions in \eqref{eq:pdf1} and \eqref{eq:pdf2}, the correlation coefficient between two reflecting elements can be derived analytically as follows
\begin{align}
\mathbf{R}_{n,m}^a=\int_{-\Delta\theta}^{+\Delta\theta}e^{jkd_{n,m}\sin(\theta_c^\text{RIS})}f_\theta(\theta_c^\text{RIS})d\theta_c^\text{RIS}.
\end{align}
In Figs. \ref{Fig:Cor}(c) and (d), we investigate the eigenvalue spread of $\mathbf{R}$ by considering the $\theta_{c,s}^\text{RIS}$ distributions given in Figs. \ref{Fig:Cor}(a) and (b), respectively.  Eigenvalue spread demonstrates a measure of the spatial correlation as well as the rank of $\mathbf{R}$. For correlated channels, non-identical eigenvalues and a low rank are obtained for $ \mathbf{R}$. In an indoor environment, since the number of strong eigenvalues are small, highly correlated channels are obtained due to the narrow range of $\theta_{c,s}^\text{RIS}$.  Meanwhile, smaller correlation coefficients are obtained in the outdoor environment, since the eigenvalues are relatively distributed equally owing to the larger spread of the scatterers. 
\begin{figure}[!t]
	\begin{center}
		\includegraphics[width=0.9\columnwidth]{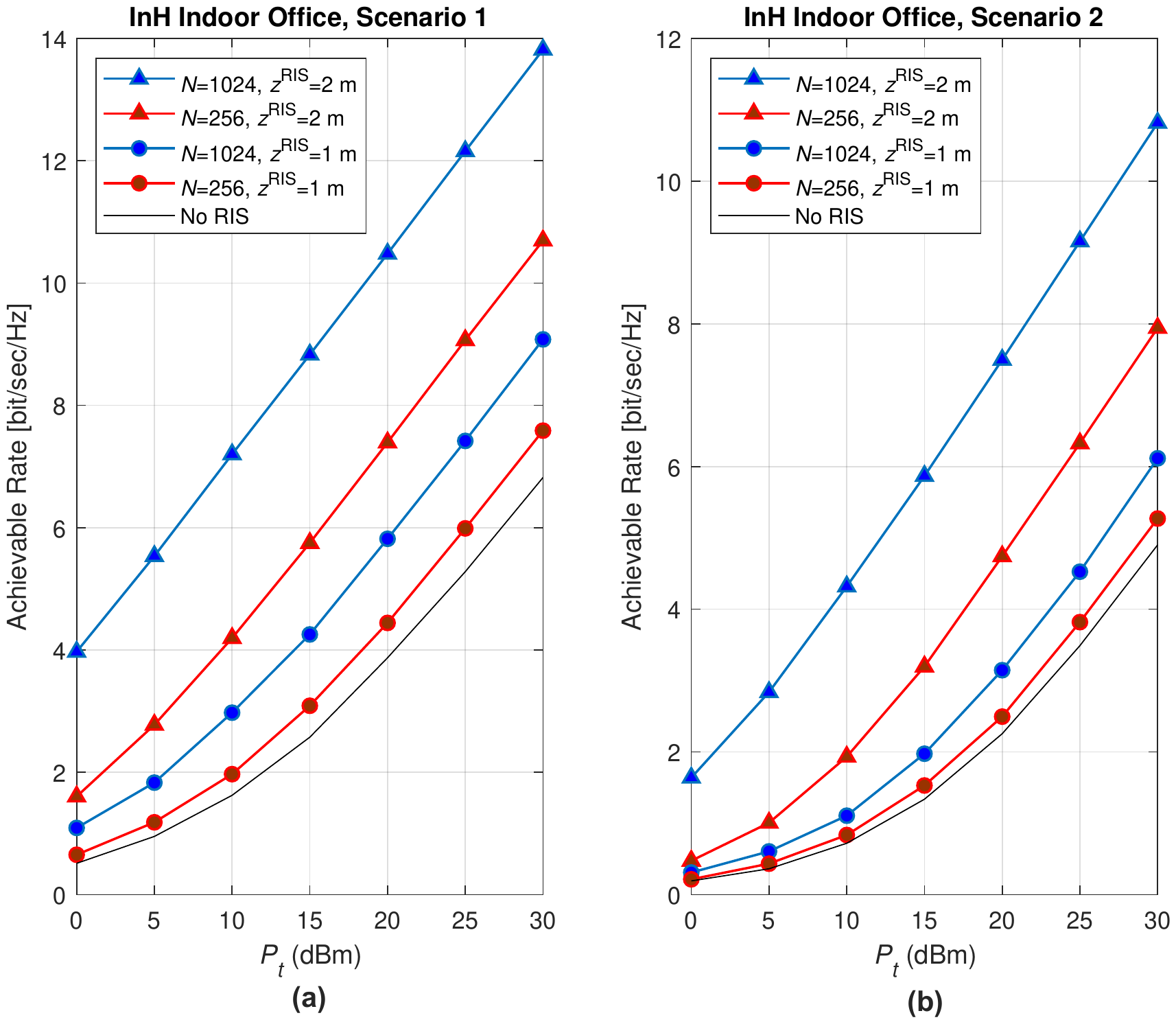}
		\caption{Achievable rates of RIS-assisted systems for indoors under (a) Scenario 1 and (b) Scenario 2.}
		\label{fig:Sim1}
	\end{center}\vspace*{-0.5cm}
\end{figure}
 Semi-analytical and analytical results are generally in agreement with simulations, while they do not exactly match with the simulation result for increasing eigenvalue numbers, since the approximations for $\theta_{c,s}^\text{RIS}$ cause small deviations for higher eigenvalue numbers in distribution as shown in Figs. \ref{Fig:Cor}(c) and (d). Nevertheless, the analytical and semi-analytical approaches give a useful insight for the characteristics of spatial correlation and motivates us to search for RIS setups with reduced channel correlation.

\section{Numerical Results}

In this section, we provide comprehensive numerical results to test our new channel model for RIS-empowered communication using the open-source SimRIS Channel Simulator MATLAB package \cite{SimRIS_latincom}. We consider the operating frequencies of $28$ and $73$ GHz, since considered system parameters are valid both these bands. The noise power is assumed to be $-100$ dBm.

\begin{figure}[!t]
	\begin{center}
		\includegraphics[width=0.9\columnwidth]{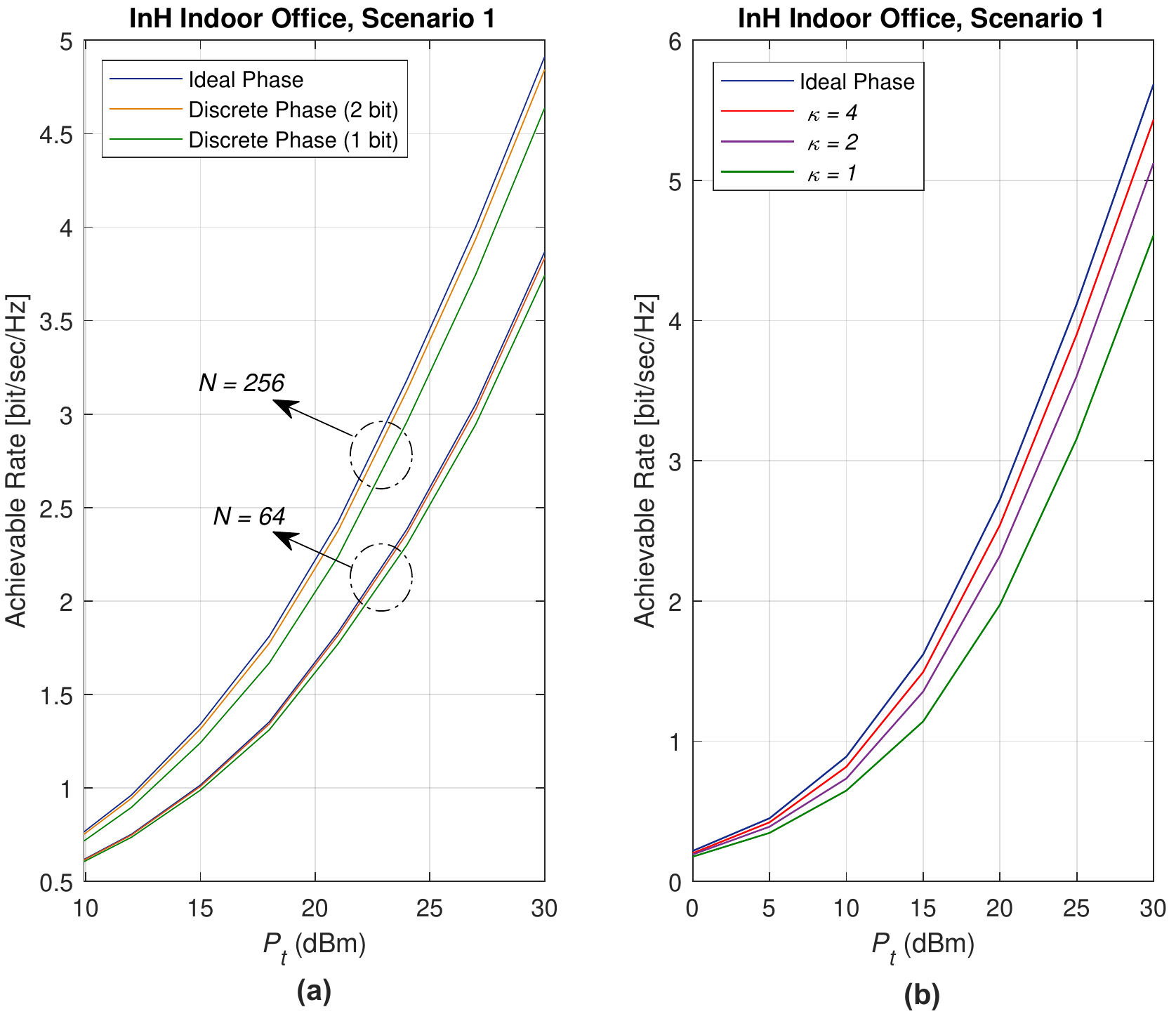}
		\caption{Achievable rates of RIS-assisted systems with (a) discrete phase shifts and (b) imperfect phase knowledge for InH Indoor Office environment.}
		\label{fig:Sim5}
	\end{center}\vspace*{-0.5cm}
\end{figure}

In Figs. \ref{fig:Sim1}(a) and (b), we evaluate the (ergodic) achievable rate $(R)$ of a communication system with and without an RIS operating in indoor environments for $N =256$ and $1024$ at $28$ GHz under the far-field conditions. Here $R$ is defined as $R=\mathrm{E}\left\lbrace \log_2(1+\rho)\right\rbrace $ [bits/s/Hz], where $\mathrm{E}\left\lbrace .\right\rbrace $ is the expectation. In Fig. \ref{fig:Sim1}(a), we consider Scenario 1 where the RIS is mounted on the side wall and the coordinates of the Tx, the Rx, and the RIS are respectively given as $(0,25,2)$, $(38,48,1)$, $(40,50,z^{\text{RIS}})$. Similarly, in Fig. \ref{fig:Sim1}(b), we consider Scenario 2 with the Tx, the Rx, and the RIS coordinates given by $(0,25,2)$, $(65,35,1)$, and $(70,30,z^{\text{RIS}})$, respectively.  Here, we consider two different RIS placements: RIS mounted at a moderate height of $z^{\text{RIS}}=1$ m and at a higher height of $z^{\text{RIS}}=2$ m. For these given parameters, we have $d_{\text{RIS-R}} \in \left\lbrace 2.82,3 \right\rbrace $ m and $d_{\text{RIS-R}} \in \left\lbrace 7.07,7.14 \right\rbrace $ m for Scenarios 1 and 2, respectively, which are valid assumptions to have a pure LOS link between the RIS and the Rx.  As seen from  Figs. \ref{fig:Sim1}(a) and (b), for the case of $z^{\text{RIS}}=1$ m, since the LOS probability is relatively low for the Tx-RIS link, which has a LOS distance of $47.1$ m, the RIS provides only a minor improvement in the received SNR. The reason of this behavior can be explained by the relatively higher attenuation of the RIS-assisted channel compared to the channel between Tx-Rx. However, a major improvement is observed for the case of $z^{\text{RIS}}=2$ m, which assumes a LOS-dominated Tx-RIS link as discussed after \eqref{eq:LOS_prob}. Finally, a noticeable degradation is observed in Fig. \ref{fig:Sim1}(b) due to larger Tx-Rx/RIS separations, which further degrades the benefits of the RIS for $z^{\text{RIS}}=1$ m. From the given results of Fig. \ref{fig:Sim1}, we conclude that the RIS can be used as an effective tool to boost the achievable rate in indoor environments when both the Tx-RIS and RIS-Rx links are LOS dominated.
\begin{figure}[!t]
	\begin{center}
		\includegraphics[width=0.85\columnwidth]{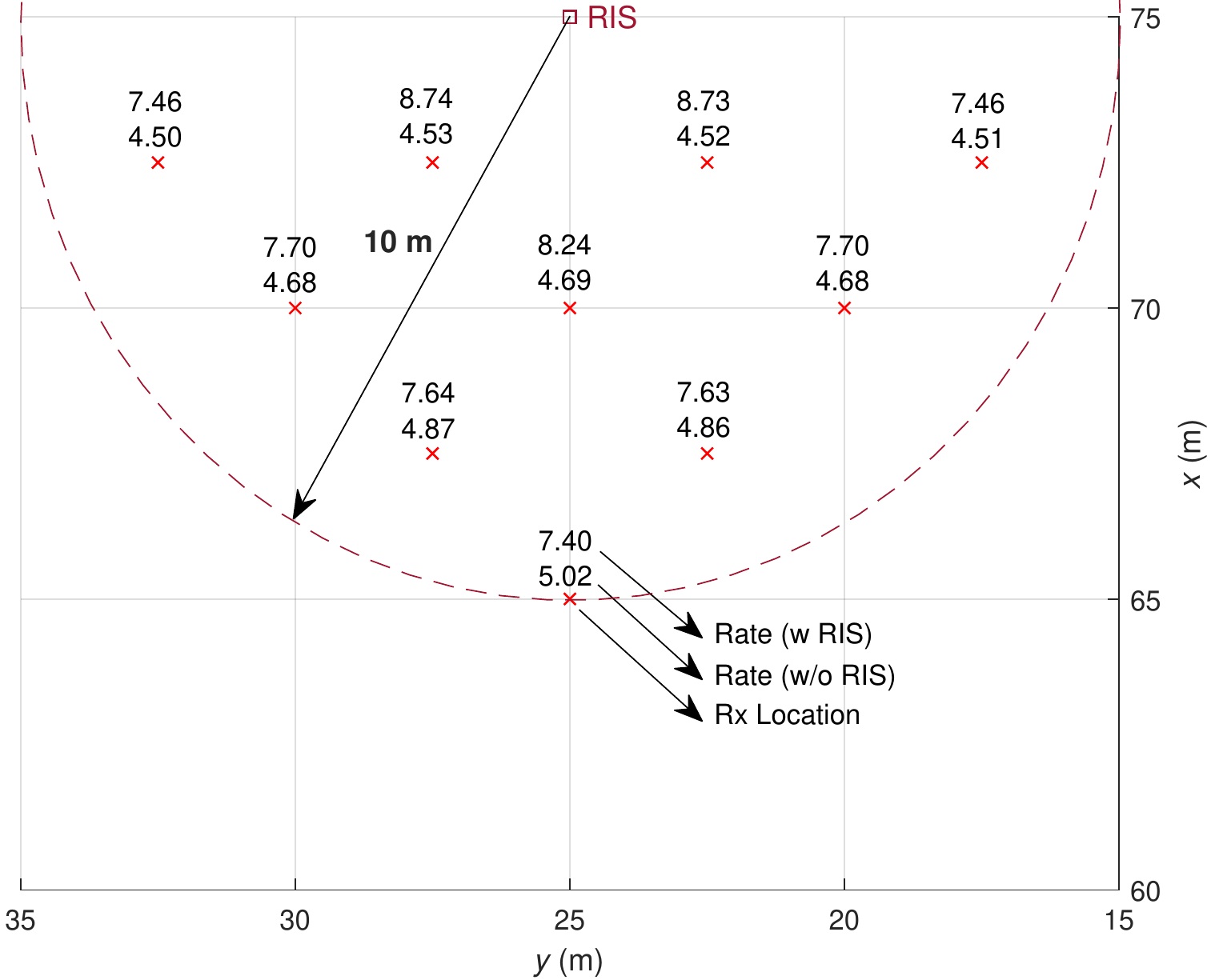}
		\vspace*{-0.3cm}\caption{Top view of the considered test scenario with $10$ reference points along with achievable rate values.}\vspace*{-0.3cm}
		\label{fig:Sim2}
	\end{center}\vspace*{-0.3cm}
\end{figure}

In Figs. \ref{fig:Sim5}(a) and (b), we investigate the effect of phase imperfections on the achievable rates of an RIS-assisted system in InH Indoor Office environment at $73$ GHz under the far-field assumption. Considering Scenario 1, the Tx and the Rx are located as in Fig. \ref{fig:Sim1}(a) while the coordinates of RIS are given by $(40,50,2)$. In Fig. \ref{fig:Sim5}(a), under $N=64$ and $256$, we compare the achievable rates of RIS-assisted systems for discrete phase shifts with $1$ and $2$ controlling bits and, ideal-continuous phase shifts over $[-\pi,\pi]$. As clearly seen, the quantization level affects the achievable rate of the system, and for increased $ N $ values, the effect of quantization errors becomes more visible. In addition, the disruptive effect of imperfect phase knowledge on the system performance is observed in Fig. \ref{fig:Sim5}(b) for $N=256$. Here, the concentration parameter $\kappa$ is inversely proportional to estimation accuracy.  Although the phase estimation and quantization errors cause degradation in achievable rate performance, the constructive effect of the RIS-controlled channel on the system remains favorable.

\begin{figure}[!t]
	\begin{center}
		\includegraphics[width=0.9\columnwidth]{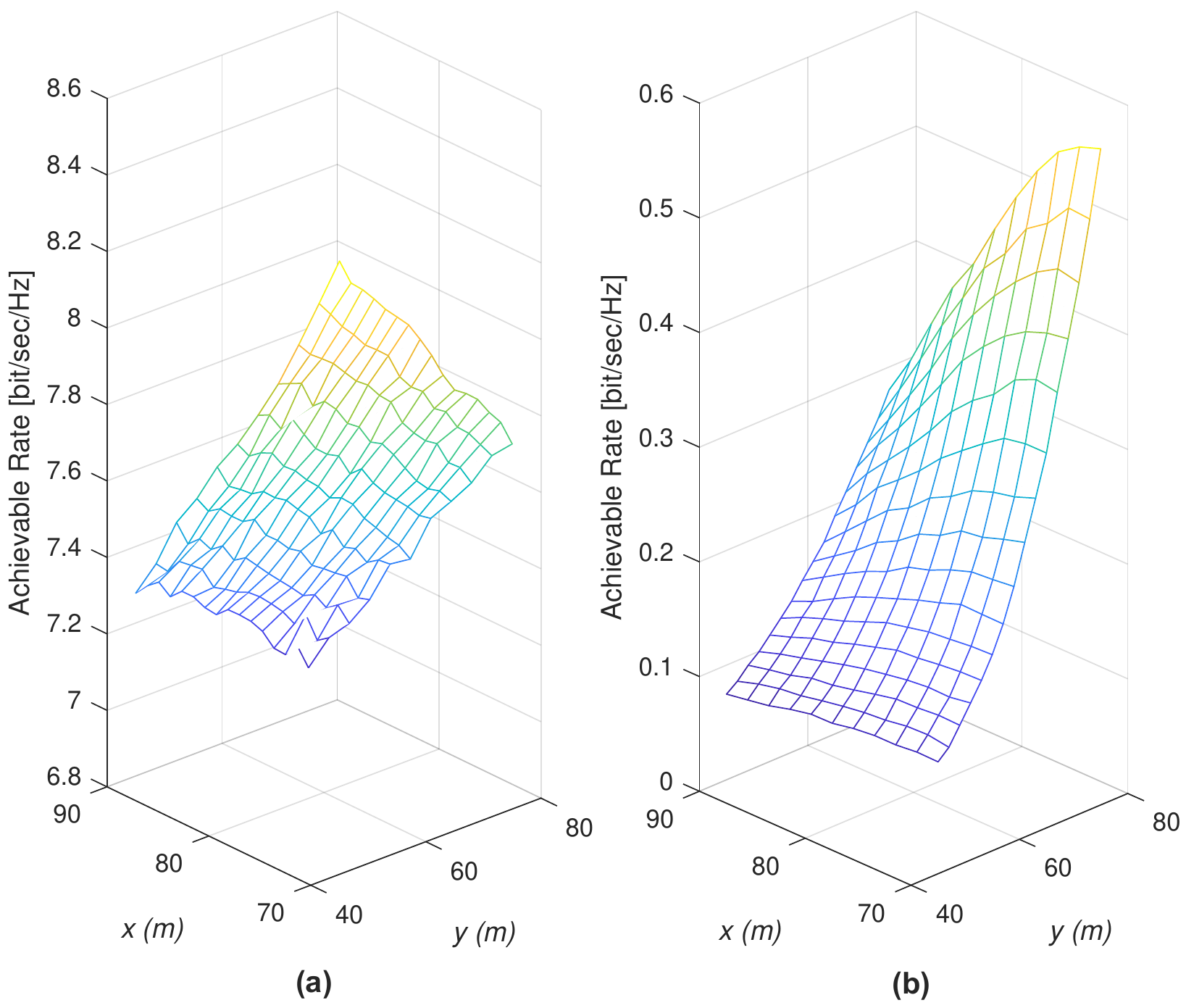}
		\vspace*{-0.2cm}\caption{Achievable rates of the an RIS-assisted system under varying Rx positions in an outdoor environment in the presence (a) and absence (b) of the direct link between the Tx-Rx.}\vspace*{-0.2cm}
		\label{fig:Sim3}
	\end{center}\vspace*{-0.2cm}
\end{figure}
In Fig. \ref{fig:Sim2}, we investigate the system achievable rate for $10$ test points within a $10$ m azimuth distance of the RIS for Scenario 2, where $N=256$ and $P_t=30$ dBm are considered. Here we assume that our LOS dominated RIS-Rx channel of $\eqref{eq:20}$ is still valid in the considered environment, which might be in the form of a open office. $(x^{\text{Rx}},y^{\text{Rx}})$ coordinates of the test points are marked on Fig. \ref{fig:Sim2}, while $z^{\text{Rx}}$ is fixed to $1$ m for all points.  The coordinates of the Tx and the RIS (Scenario 2) are given respectively as $(0,25,2)$ and $(75,25,2)$, where the distances between the RIS and test points vary between $\sqrt{13.5}$ and $\sqrt{101}$ m. In this setup, the RIS has a clear LOS path with the Tx, and reflects the incoming signals in an effective way to boost the rate. As seen from Fig. \ref{fig:Sim2}, a significant improvement is obtained in $R$ with an RIS, particularly for the test points closer to the RIS. We observe that even at a $10$ m azimuth distance from the RIS, around $1.6$ bits/s/Hz improvement can be provided by an RIS. We further observe that $R$ is quite sensitive to the length of the RIS-Rx link by varying up to $1.35$ bits/s/Hz and might be identical for certain points due to our symmetrical setup (aligned Tx-RIS link).

In Figs. \ref{fig:Sim3}(a) and \ref{fig:Sim3}(b), the effect of varying Rx positions on the achievable rate of an RIS-assisted system at $28$ GHz is examined for a UMi Street Canyon outdoor environment. 
Here, we consider Scenario 1 where the RIS is mounted on the side wall and the coordinates of the Tx, the Rx, and the RIS are respectively given as $(0,25,20)$, $(x^{\text{Rx}},y^{\text{Rx}},1)$, $(70,85,10)$. In Fig. \ref{fig:Sim3}(a), the direct link between the Tx-Rx is available as well as the RIS-assisted link for transmission, while it is assumed that the direct link between Tx-Rx is blocked in Fig. \ref{fig:Sim3}(b). In Fig. \ref{fig:Sim3}(a), we observe that the highest achievable rate is obtained where the Rx is close to the Tx, since the channel between Tx-Rx is more dominant than the RIS-assisted link in terms of achievable rate. Furthermore, if the Rx moves to an area far from the RIS on $ x $ and $y$-axes, the effect of the RIS will drastically diminish as shown in Fig. \ref{fig:Sim3}(b). We also observe that the most critical system parameter is the separation between RIS-Rx when the direct link between Tx-Rx is blocked.

\begin{figure}[!t]
	\begin{center}
		\includegraphics[width=0.8\columnwidth]{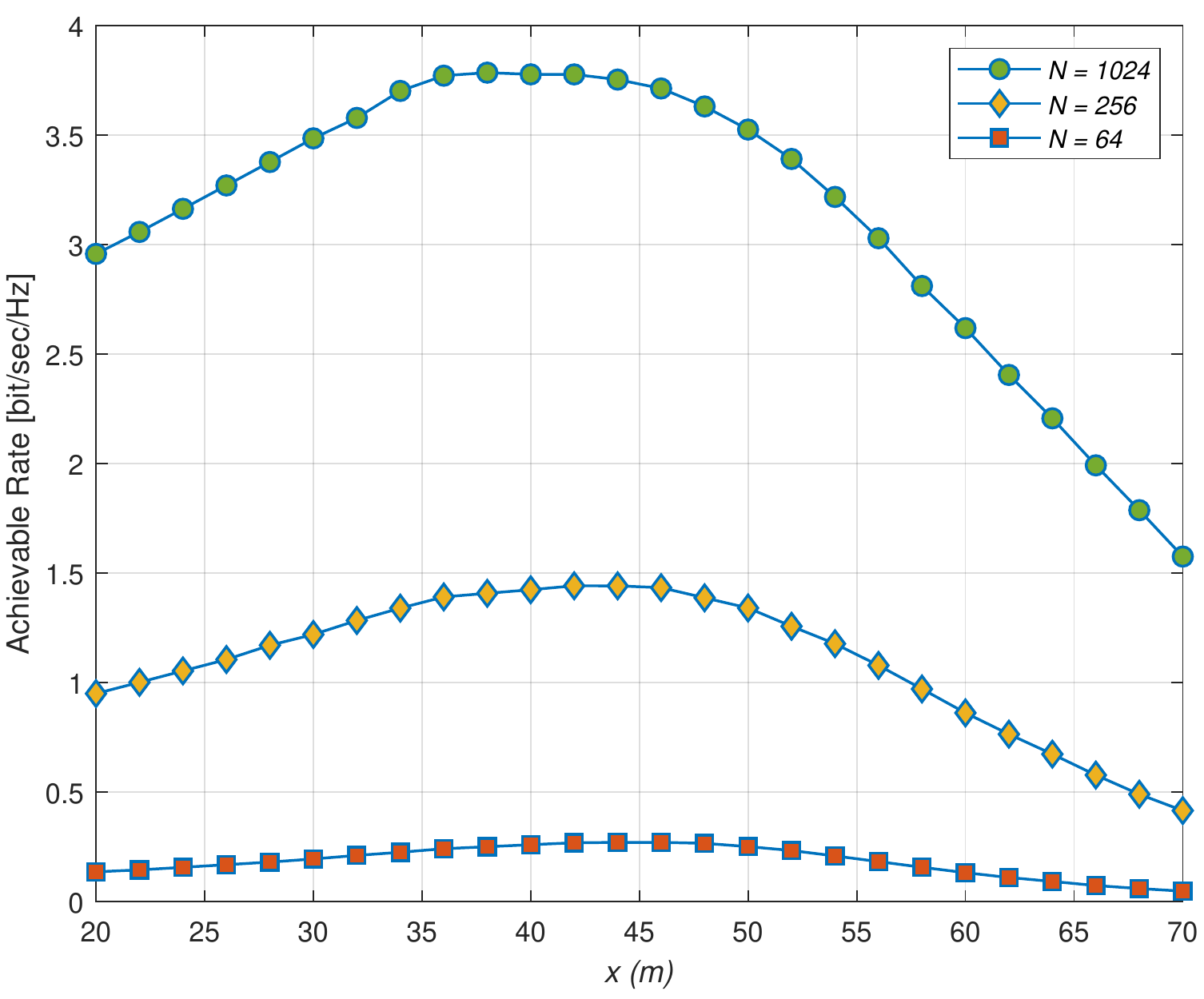}
		\vspace*{-0.2cm}\caption{Achievable rate of the an RIS-assisted system under varying RIS positions in an outdoor environment.}\vspace*{-0.3cm}
		\label{fig:Sim6}
	\end{center}
\end{figure}

In Fig. \ref{fig:Sim6}, the effect of the RIS position on the achievable rate is examined in the absence of a direct path between Tx-Rx in an outdoor environment for $N=64$, $256$ and $1024$. Considering Scenario 1 and UMi Street Canyon model at $28$ GHz, the coordinates of the Tx, the Rx, and the RIS are respectively given as $(0,25,20)$, $(50,50,1)$, $(x^{\text{RIS}},60,10)$. As clearly seen in Fig. \ref{fig:Sim6}, a noticeable increase in achievable rate is obtained at regions where the RIS is closely aligned with the Rx on the $ x $-axis. Moreover, when the RIS moves away from the Rx and the Tx, a substantial decrease in achievable rate is observed and reliable transmission cannot be guaranteed. We conclude that in outdoors, the most effective scenario is obtained when $d_{\text{RIS-R}}$ is kept less than $ 15 $ m.

Another important scenario that should be considered when designing an RIS-assisted system is the placement of the RIS in the near-field of the Rx. If $N$ increases considerably, the far-field assumption is no longer valid and near-field conditions should be considered to investigate the performance. Placing the RIS in the near-field of the Rx will lead to a pure LOS link between the RIS and Rx. In Fig. 11, we evaluate the achievable rate of an RIS-assisted communication system operating in indoor environments for varying $N$ values at $28$ GHz under near-field conditions. Here, the channel gain for the RIS-Rx link is approximated as in \cite{Emil_power}. Moreover, we consider Scenario 1 where the RIS is mounted on the side wall, and the coordinates of the Tx, the Rx, and the RIS are respectively given as ($0,25,2$), ($55,35,1$), ($50,30,2$). As clearly seen, increasing $N$ values provide a significant improvement in the achievable rate even in the near-field conditions. As seen from Fig. 11, a decrease of approximately $10$ dB is obtained in $P_t$ when the $N$ value is quadrupled, while a decrease of approximately $15$ dB is obtained in the far-field case as seen from Fig. 6(a). From the given results, we conclude that the effect of increasing $N$ value in the far-field conditions is more dominant than the near-field conditions, and the provision of far-field conditions will be beneficial when designing the RIS-assisted communication system at mmWave frequencies.

\section{Conclusions}

This paper has been a first step towards physical channel modeling with RIS-empowered networks and aimed to create a new line of research for wireless researchers.  Our SimRIS Channel Simulator package can be used effectively in Monte Carlo simulations to assess the capacity, SNR gain, secrecy, outage and error performance of RIS-assisted systems by providing cascaded channel coefficients separately. We also note that our channel modeling methodology can be easily extended for multi-carrier cases by considering time delays of each particular path. Our future research questions are summarized below: \textit{i)} What about the delay spread? Can an RIS reduce the delay spread in wideband channel models? \textit{ii)} What is the statistical distribution of the composite channel ($h$)? While we expect it to follow Gaussian distribution due to the Central Limit Theorem, what is the influence of an RIS? \textit{iii)} How fast $h$ changes  and what effect will this have on correlation? Can the RIS slow down (or speed up) the rate of change of $h$?  \textit{v)} How realistic? Extensive measurements are required to obtain more realistic system parameters for RIS-assisted links.

\begin{figure}[!t]
	\begin{center}
		\includegraphics[width=0.86\columnwidth]{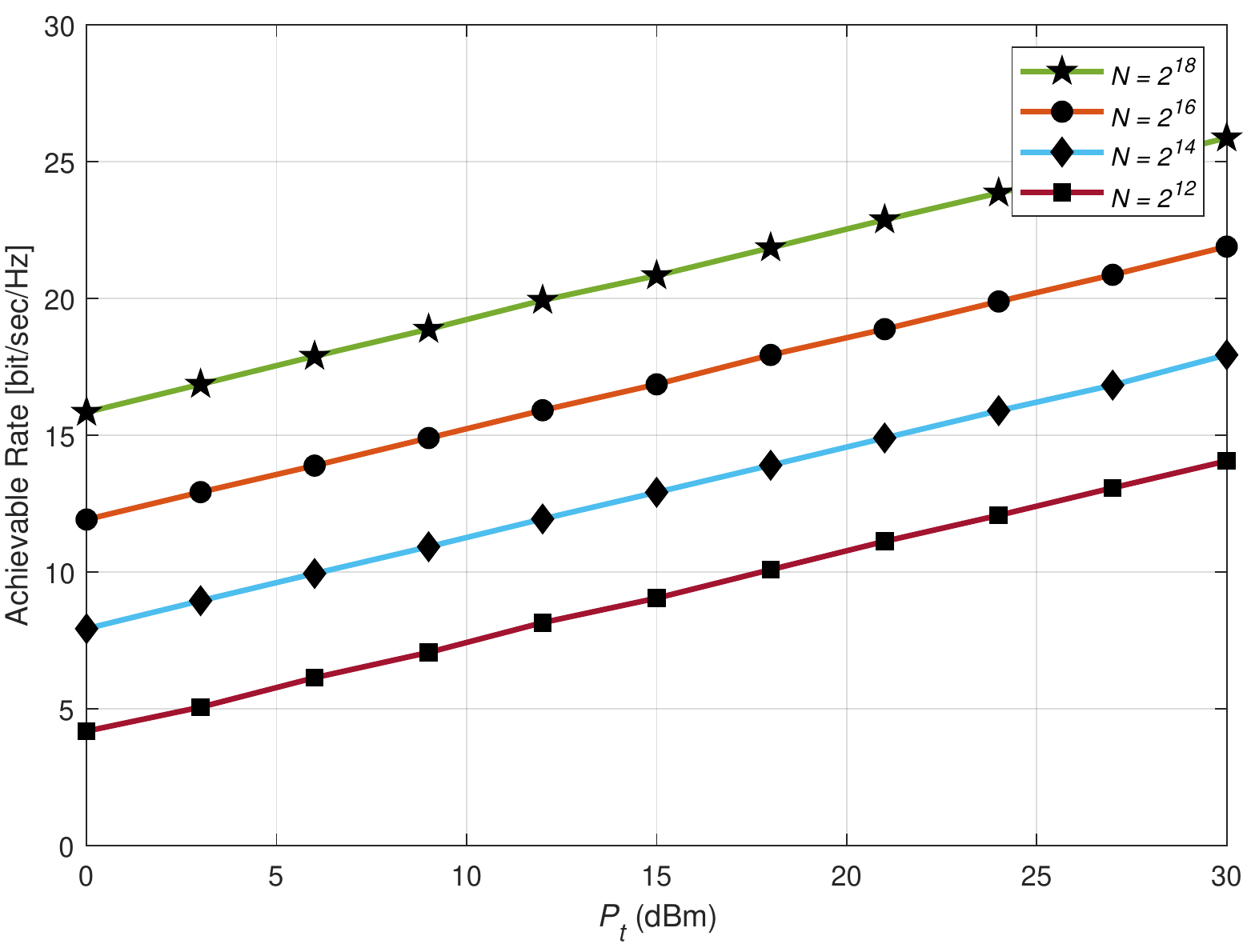}
		\vspace*{-0.2cm}\caption{Achievable rate of the an RIS-assisted system under the near-field assumption in an indoor environment.}\vspace*{-0.3cm}
		\label{fig:Sim_Near}
	\end{center}
\end{figure}

\bibliographystyle{IEEEtran}
\bibliography{bib_2020}

%
%
%
\begin{IEEEbiography}[{\includegraphics[width=1in,height=1.25in,clip,keepaspectratio]{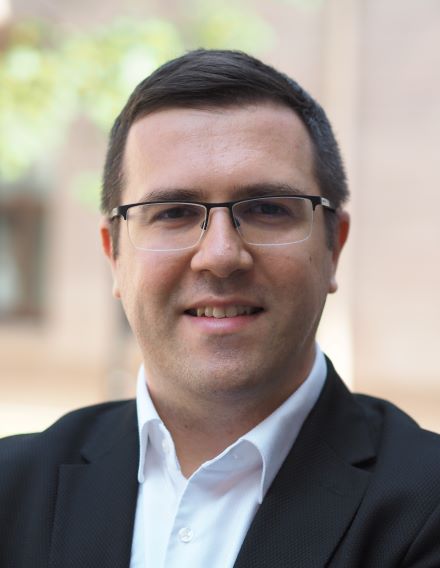}}]{Ertugrul Basar} received his Ph.D. degree from Istanbul Technical University in 2013. He is currently an Associate Professor with the Department of Electrical and Electronics Engineering, Ko\c{c} University, Istanbul, Turkey and the director of Communications Research and Innovation Laboratory (CoreLab). His primary research interests include beyond 5G systems, index modulation, intelligent surfaces, waveform design, and signal processing for communications. Dr. Basar currently serves as a Senior Editor of \textsc{IEEE Communications Letters} and an Editor of \textsc{IEEE Transactions on Communications} and \textit{Frontiers in Communications and Networks}. He is a Young Member of Turkish Academy of Sciences and a Senior Member of IEEE.

\end{IEEEbiography}

\begin{IEEEbiography}
	[{\includegraphics[width=1in,height=1.25in,clip,keepaspectratio]{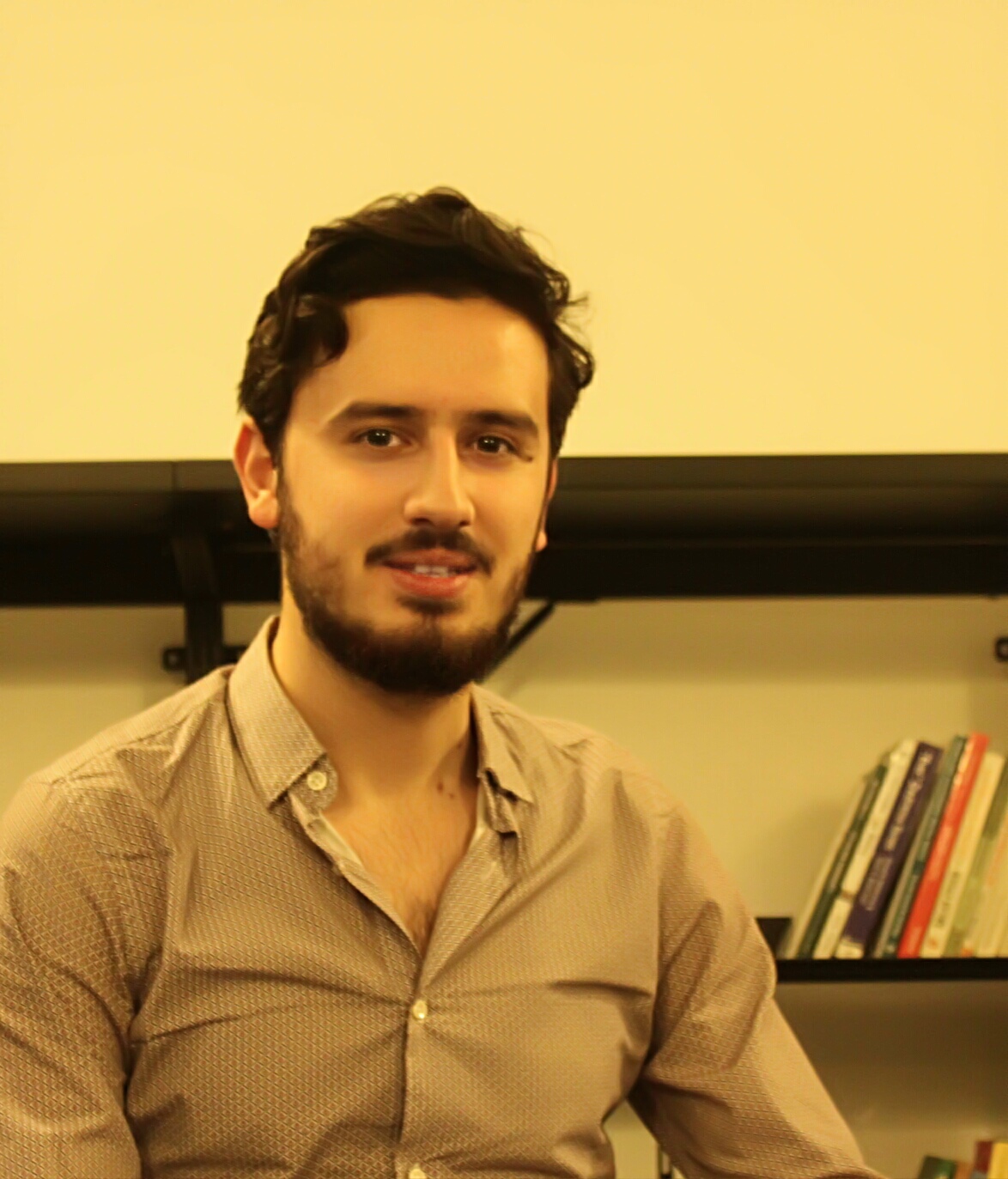}}]{Ibrahim Yildirim} received his B.S. and M.S. degrees from Istanbul Technical University, Turkey, in 2017 and 2019, respectively. He is currently pursuing the Ph.D. degree at Ko\c{c} University. He is also a Research and Teaching Assistant at Istanbul Technical University. His current research interests include MIMO systems and reconfigurable intelligent surfaces. He has been serving as a Reviewer for  \textsc{IEEE Journal on Selected Areas in Communications}, \textsc{IEEE Transactions on Vehicular Technology}, and \textsc{IEEE Communications Letters}.
	
\end{IEEEbiography}
\begin{IEEEbiography}
	[{\includegraphics[width=1in,height=1.25in,clip,keepaspectratio]{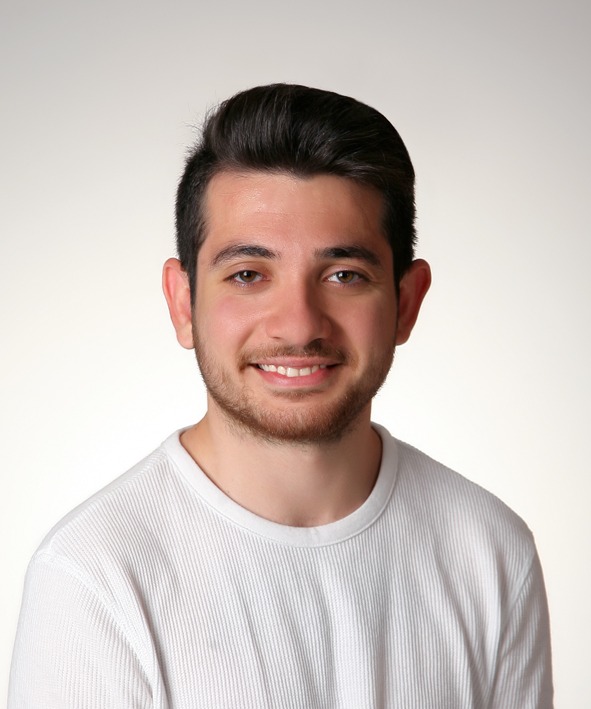}}]{Fatih Kilinc} received his B.S degree from Istanbul Medipol University in 2020. He is currently pursuing M.S. degree at Ko\c{c} University. He is a research and teaching assistant at Ko\c{c} University. His research interest include channel modeling, intelligent surfaces and signal processing for wireless communications.
	
\end{IEEEbiography}

\end{document}